\documentclass{aa}  

\usepackage{graphicx}
\usepackage{txfonts}
\usepackage{lipsum}
\usepackage{subcaption}         % necessary for continued figures, example in section 3
\usepackage{lscape}             % to rotate a single page table, example in appendix.
\usepackage{placeins}           % useful with \FloatBarrier, to keep 
\usepackage{hyperref}

\usepackage[normalem]{ulem}
\usepackage{amsmath}
\usepackage{color} %\definecolor{purple}{rgb}{0.5,0,0.87}

\newcommand\dm{{\rm DM}}
\newcommand\df{DF}

\newcommand{\be}{\begin{equation}}
\newcommand{\ee}{\end{equation}}
\newcommand{\ben}{\begin{eqnarray}}
\newcommand{\een}{\end{eqnarray}}

\begin{document}

\title{
The slow evolution of dark matter halos from cusp to core naturally produces extended stellar core-like distributions
}

\author{Jorge S\'anchez Almeida\inst{1,2}
  \and
  Angel R. Plastino\inst{3}
  \and
   Ignacio Trujillo\inst{1,2}
       }
%%%
       \institute{
       {Instituto de Astrof\'\i sica de Canarias, La Laguna, Tenerife, E-38200, Spain}\\
       \email{jos@iac.es,itc@iac.es}
       \and
       Departamento de Astrof\'\i sica, Universidad de La Laguna
       \and
       {CeBio y Departamento de Ciencias B\'asicas, 
       Universidad Nacional del Noroeste de la Prov. de Buenos Aires, 
       UNNOBA, CONICET, Roque Saenz Pe\~na 456, Junin, Argentina}\\
       \email{arplastino@unnoba.edu.ar}
       }
   \date{Received \today; accepted \dots}
  \abstract{
Motivated by the observation of extended stellar cores in dark matter (DM) dominated dwarf galaxies, this study investigates a simple mechanism by which stellar cores can form as a result of DM halo expansion. Several non-CDM models predict that the DM distribution thermalizes over time, transforming initially cuspy halos into cores. This transformation weakens the gravitational potential, allowing the stellar component to expand and form diffuse, core-like structures. Using analytical models and adiabatic invariants, we examine stellar systems with purely tangential, purely radial, and isotropic orbits evolving under a slowly changing potential. Across a wide range of initial and final conditions, we find that stellar cores form relatively easily, though their properties depend sensitively on these conditions. Orbit types preserve their nature during the \dm\  halo expansion: tangential and radial orbits remain so, while isotropic orbits remain nearly isotropic in the central regions. Systems with circular orbits develop stellar cores when the initial stellar density  logarithmic  slope lies between $-0.5$ and $-1.2$, whereas radial systems do not form cores. Isotropic systems behave similarly to tangential ones, producing cores that are isotropic in the center but develop increasing radial anisotropy outward; the anisotropy parameter $\beta$ grows from ~0.07 at the core radius to $\sim$0.5 at three core radii. The theoretical and observational literature suggests initial DM profiles with steep slopes and stellar distributions that are shallower and isotropic at the center. Given these conditions, the mechanism predicts stellar cores with radii at least 40\% that of the DM core and inner logarithmic slopes shallower than 0.6. 
}
\keywords{
  Galaxies: dwarf --
  Galaxies: evolution --
  Galaxies: halos --
  Galaxies: kinematics and dynamics --
  Galaxies: stellar content --
  dark matter %--
}
\titlerunning{Gradual halo core formation produces extended stellar cores}
\authorrunning{S\'anchez Almeida et al.}
\maketitle

%\graphicspath{{../figures+/}}


%%%%%
\section{Introduction}\label{sec:intro}
Although known from the 80's \citep{1984AJ.....89...64B,1987AJ.....94.1126C}, a new generation of deep surveys are disclosing a large number of faint low surface brightness galaxies with extended stellar distributions. They receive several names depending on the technique of observation and galaxy mass, including ultra diffuse galaxies \citep[UDG; e.g.,][]{2015ApJ...798L..45V,2017MNRAS.468.4039R,2018RNAAS...2...43C,2018A&A...615A.105M,2021NatAs...5.1182T} or ultra faint galaxies \citep[UFD; e.g.,][]{2019ARA&A..57..375S,2024ApJ...967...72R,2025ApJ...982L...3A}. They tend to be satellites of nearby larger galaxies, but this is probably an observational bias given the difficulties to detect faint low-surface brightness sources at large distances in low density fields.  One of their more conspicuous features is the lack of clear concentration of stars marking the center of the stellar distribution not obvious at first sight. An extreme case is the galaxy Nube \citep{2024A&A...681A..15M}, with a large half-mass radius (6.9\,kpc) and a mass distribution  with a central plateau of low mass surface density ($\sim1\,M_\odot\,{\rm pc}^{-2}$), but there are many other examples \citep{2015ApJ...809L..21M,2022ApJ...933...47C,2025ApJ...982L...3A}.
According to the standard picture, these spread-out stellar distributions could be caused by stellar feedback in low-mass dark matter halos \citep[e.g.,][]{2017MNRAS.466L...1D,2017MNRAS.469.2335C,2020MNRAS.497.2393L}. However, the presence of these stellar density plateaus or cores have been detected in objects too small for the stellar feedback mechanism to operate \cite[][]{2024ApJ...973L..15S}. The question arises as whether mechanisms other than the stellar feedback can also produce the observed extended stellar cores.    

Here we analyze a conceptually simple physical mechanism to produce spread-out stellar cores in \dm\ dominated halos with stars. If internal processes make the DM evolve with time to create a core, then the gravitational force that maintains the stellar system drops. The stellar system expands dragged along by the expansion of \dm\ creating a diffuse system of stars with a core. A number of physical models of \dm\  going beyond C\dm\ produce the described slow expansion; for instance, self-interacting \dm\ \citep[SIDM;][]{2000PhRvL..84.3760S} turns cusps into cores within a timescale of the order of the Hubble time \citep[$t_H$;  e.g.,][]{2023MNRAS.523.4786O}. The timescale for the evolution of warm \dm\ halos is similar to the formation of C\dm\ halos, and so, also around $t_H$ \citep{2000ApJ...542..622C,2001ApJ...556...93B}. Fuzzy \dm\ \citep{2014NatPh..10..496S}, self-interacting fuzzy \dm\ \citep{2025arXiv250204838I}, and Bose-Einstein condensates \citep{2023MNRAS.518.4064D} all lead to the formation of cores on timescales shorter than in SI\dm , but still longer than the dynamical timescale \citep[e.g.,][]{2017MNRAS.471.4559M}. Fermionic \dm\ evolves on cosmological timescales  \citep{2023Univ....9..197A}, whereas late-time \dm\ decay \citep{2018JCAP...07..013C} leads to the gradual formation of \dm\ cores at late times. These are just illustrative examples, however, we stress that the mechanism is independent of the actual \dm\ model ultimately responsible for the expansion.  In a sense, it is inverse to the adiabatic contraction suffered by C\dm\ halos early on, when the sinking of baryons towards the center of the potential drives the compaction of the \dm\ halos \citep{1986ApJ...301...27B,2004ApJ...616...16G}. In our case the expansion of the \dm\ halo drives the expansion of the stellar system.

Our work aims at showing that the hypothetical mechanism seems to work in practice.  We use a number of simplified spherically-symmetric analytical models showing that the expansion of almost any DM halo giving rise to a core  drags along the pre-existing stars often producing a core-like structure in the inner stellar distribution. Depending of the initial conditions, this core is comparable in size with the DM core, a fact consistent with the observations of some early-type dwarf galaxies having the two radii available \citep[][]{2024A&A...690A.151S,2024ApJ...973L..15S,2025A&A...694A.283S}.%\comment{Nacho:complete references}

Since the underlying assumption is that variations in the DM distribution occur slowly, on a timescale much longer than the period of stellar orbits, action variables remain constant, simplifying the study of the process \citep[e.g.,][]{2008gady.book.....B}. The final stellar distribution depends only on the initial and final distribution of \dm\ and on the initial distribution of stars, being independent of the actual pathway from initial to final. 

The paper is organized as follows:
Sect.~\ref{sec:models} describes three types of analytical spherically-symmetric systems with different and extreme velocity anisotropies covering all possibilities, from purely tangential to purely radial orbits.
Section~\ref{sec:spherical} treats models with circular orbits that have only tangential velocities, characterized by a  velocity anisotropy parameter $\beta=-\infty$. ($\beta$ is defined in Eq.~[\ref{eq:defbeta}].)
In general, the action variables are not analytical, however, the Henon's isochrone potential \citep{1959AnAp...22..126H,2008gady.book.....B} is specially useful in our context because the action variables can be expressed in closed analytical form. We use it  in Sect.~\ref{sec:isochrone} to study initially isotropic stellar orbits  ($\beta = 0$). Finally,  purely radial orbits in are considered in Sect.~\ref{sec:radial} ($\beta=1$). (Appendix~\ref{sec:general} also show how numerically handle arbitrary potentials.) We discuss the behavior under a broad range of initial and final conditions, finding that stellar cores are easily produced for a wide range of initial conditions. However, the properties of the induced stellar cores depend on the initial conditions. Thus, in Sect.~\ref{sec:literature} we revise the existing literature on the initial conditions to be expected, including the \dm\ and stellar mass density distributions as well as the velocity anisotropy of the stars. Appendix~\ref{app:blablabla} analyzes the expected relative half-mass radii of \dm\ and stars when halos first formed in the early Universe. The constraints set by the expected initial conditions on the resulting stellar core-like structures are discussed in Sect.~\ref{sec:discussion}. Finally, the results and conclusions to be extracted from our simple modeling are summarized in Sect.~\ref{sec:conclusions}.
All the analysis neglects the self-gravity of the stars, which is well justified for some objects \citep[e.g.,][]{2024ApJ...973L..15S} but not in general. The impact of the self-gravity of the stars is briefly addressed in Sect.~\ref{sec:conclusions}.
Appendix~\ref{sec:star_trace_dm} addresses why the relative expansion of the stellar cores induced by the mechanism agrees with the expansion of the \dm\ core.

%%%%%%%
%
\section{Analytical models to treat the adiabatic expansion of the stellar system}\label{sec:models}
We consider three types of spherically symmetric systems with different and extreme velocity anisotropies covering all possibilities. As it is mentioned in the introduction (Sect.~\ref{sec:intro}), the assumed slow variation in the \dm\ distribution makes it possible to describe the time evolution using adiabatic invariants, which remain conserved throughout the process. The contribution of the stars to the potential is in all cases assumed to be negligible. Thus, our modeling is meant to describe dwarf galaxies fully dominated by DM at all times.  The main difference between models is the anisotropy of the stellar orbits with are usually described using the anisotropy parameter, $\beta$, defined as,
\begin{equation}
  \beta = 1 -\frac{\sigma_t^2}{2\sigma_r^2},
  \label{eq:defbeta}
\end{equation}
with $\sigma_t^2$ the dispersion of tangental velocities and $\sigma_r^2$ the dispersion of radial velocities at a given radial distance \citep[e.g.,][]{2008gady.book.....B}. Stellar systems with circular orbits have $\sigma_r^2=0$ and so are characterized by $\beta=-\infty$, systems with isotropic orbits have $\beta=0$, and systems with pure radial orbits have $\beta = 1$. They are treated in Sects.~\ref{sec:spherical}, \ref{sec:isochrone}, and \ref{sec:radial}, respectively. The general case of arbitrary potential and $\beta$ is put forward in Appendix~\ref{sec:general}. As it is mentioned above, the self-gravity of the stars is neglected, but the impact of the self-gravity of the stars is briefly considered in Sect.~\ref{sec:conclusions}.
%
%%%%%%%%
%
\subsection{Circular stellar orbits in arbitrary spherical potentials}\label{sec:spherical}

Here, we consider spherically symmetric stellar systems with purely tangential orbits living in an externally imposed potential which is time dependent but always spherical.  Since the potential is spherical and there are no initial radial velocities, the initial orbits are circular and depend only on the radial coordinate $r$. In this idealized situation, the gravitational forces are perfectly balanced with the centrifugal forces so that,
\begin{equation}
  v^2_\star(r,t)=G\,M_{\dm}(<r,t)/r,
\label{eq:balance1}
\end{equation}
with $v_\star$ the circular velocity of the stars and $M_{\dm}(<r,t)$ the \dm\ mass  internal to $r$,
\begin{equation}
  M_\dm(<r,t) = 4\pi\, \int_0^r\rho_\dm(r',t)\,r'^2dr',
  \label{eq:innermass}
\end{equation}
which is assumed to be distributed with a volume density $\rho_\dm(r,t)$. As usual, $G$ stands for the gravitational constant.
In the case of circular orbits, the radial invariant is zero (Eq.[\ref{numeJr}], with $r_{\min} = r_{\max} = r$). Since the adiabatic invariants are conserved, it is also zero in the final orbits, so that the initial circular orbits remain circular, and the balance given by Eq.(\ref{eq:balance1}) is maintained at any time $t$.
The angular momentum of the orbits is also conserved, so that the plane of each orbit remains invariant. Thus, a star starting at $t_0$ with an orbit of radius $r_0$ and ending at $t$ with an orbit of radius $r$ has to comply with the condition 
\begin{equation}
  r\,v_\star(r,t) = r_0\,v_\star(r_0,t_0).
  \label{eq:angular_momentum}
\end{equation}
The stellar mass is also conserved during expansion, therefore, if a stellar shell at $t_0$ becomes another shell at $t$ then
\begin{equation}
  4\pi r^2\rho_\star(r,t)\, \Delta r= 4\pi r_0^2\rho_\star(r_0,t_0)\, \Delta r_0,
 \label{eq:mass_conserv}
\end{equation}
with $\rho_\star(r,t)$ the density in the shell at $r$ and $t$ and  $\Delta r_0$ and $\Delta r$ the widths of the initial and final shells.
Given Eqs.~(\ref{eq:balance1}) and (\ref{eq:angular_momentum}), the radii of the orbits have to scale as
\begin{equation}
  r=r_0\frac{M_\dm(<r_0,t_0)}{M_\dm(<r, t)},
  \label{eq:scaling}
\end{equation}
which also sets the ratio between the width of the shells used in Eq.~(\ref{eq:mass_conserv}),
\begin{multline}
  \Delta r \left[M_\dm(<r,t)+4\pi r^3\rho_\dm(r,t)\right] = \\
  \Delta r_0 \left[M_\dm(<r_0,t_0)+4\pi r_0^3\rho_\dm(r_0,t_0)\right].
  \label{eq:deltas}
\end{multline}
Equations~(\ref{eq:mass_conserv}), (\ref{eq:scaling}), and (\ref{eq:deltas}) provide the relationship between the initial distribution of stars at $t_0$ and the final one,
\begin{multline}
  \rho_\star(r,t) = \rho_\star(r_0,t_0)\times \left[\frac{M_\dm(<r,t)}{M_\dm(<r_0,t_0)}\right]^2\times\\
                                                                                                                \left[\frac{M_\dm(<r,t)+4\pi r^3\rho_\dm(r,t)}{M_\dm(<r_0,t_0)+4\pi r_0^3\rho_\dm(r_0,t_0)}\right],
\label{eq:master}
\end{multline}
once the initial and final distribution of \dm\ are set. Specific examples are given in Sect.~\ref{sec:example_circular}. 
%
%%%
\subsubsection{Densities following power laws}\label{sec:power_laws}

The inner parts of the density profiles used to represent \dm\ and stellar distributions are often power laws, including the NFW profile \citep{1997ApJ...490..493N}  characteristic of C\dm\ halos,  the polytropes representing self-gravitating systems of maximum entropy \citep{1993PhLA..174..384P,2020A&A...642L..14S}, the early \dm\ halos (see Sect.~\ref{sec:literature}), as well the $abc$ profiles commonly used to model mass density profiles in galaxies \citep[][]{1990ApJ...356..359H,2006AJ....132.2685M},
\begin{equation}
  \rho(r)=\frac{\rho_s}{(r/r_s)^c\,\left[1+(r/r_s)^a\right]^{(b-c)/a}}.
  \label{eq:abcprofile}
\end{equation}
This $abc$ profile is  defined by a characteristic density and radius ($\rho_s$ and $r_s$, respectively) and three exponents ($a$, $b$, and $c$).  The coefficient $c$ sets the inner logarithmic slope of the profile since
\begin{equation}
  \rho(r)\propto r^{-c} {\rm ~~~~when~} r\to 0.
\end{equation}
If the initial and final \dm\ profiles as well as the initial stellar distribution are of this kind, then the final stellar profile is also a power law whose  slope is  
\begin{equation}
  c_\star(t) = c_\star(t_0)\frac{4-c_\dm(t)}{4-c_\dm(t_0)}+3\frac{c_\dm(t)-c_{\dm}(t_0)}{4-c_\dm(t_0)}.
  \label{eq:slopes}
\end{equation}
The previous expression considers that the inner slope of the \dm ,  $c_\dm(t)$, and the stars, $c_\star(t)$, vary with time, and it can be derived from   Eqs.~(\ref{eq:innermass}) and (\ref{eq:master}). In the case of interest, the \dm \ profile ends up with a core, i.e.,  $c_\dm(t)=0$. Then the stars also end up with an inner core, $c_\star(t)=0$,  when
\begin{equation}
  c_\star(t_0) = \frac{3}{4} c_\dm(t_0).
  \label{eq:constraint1}
\end{equation}
Note also that according to Eq.~(\ref{eq:slopes}), the final slope of the stars may be positive, $c_\star(t) < 0$, provided the difference of initial inner slopes between stars and dark matter is large enough,
\begin{equation}
  c_\star(t_0) < \frac{3}{4} c_\dm(t_0).
  \label{eq:other_constraint}
\end{equation}
Positive final stellar slope implies the possibility of generating an inner depresion in the stellar distribution due to the expansion of the DM halo. In case the initial \dm\ and stellar profiles have the same inner slope, $c_\star(t_0)=c_\dm(t_0)$, then the final stellar slope is positive but much smaller than the original one,
\begin{equation}
 c_\star(t)\simeq \frac{c_\dm(t_0)}{4-c_\dm(t_0)}.
\end{equation}
In the case of an initial NFW distributions, $c_\dm(t_0)=1$  and therefore $c_\star(t)=1/3$.
All the above properties are illustrated in Sect.~\ref{sec:example_circular}. 
%
%%%
\subsubsection{Examples of systems with circular orbits}\label{sec:example_circular}
\begin{figure*}[ht!] 
\centering
\includegraphics[width=0.33\linewidth]{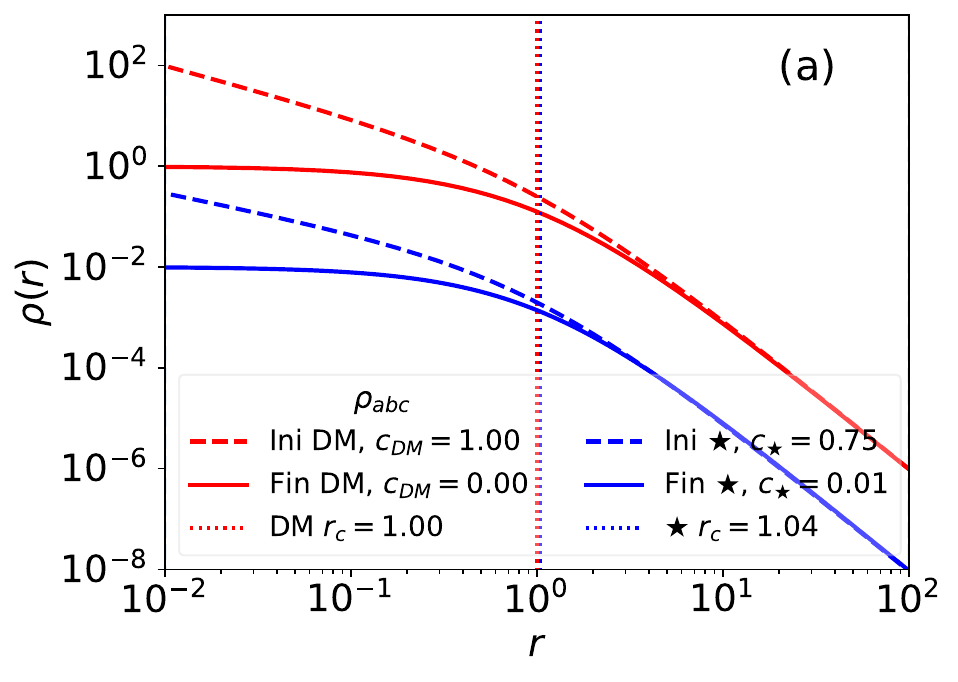}
\includegraphics[width=0.33\linewidth]{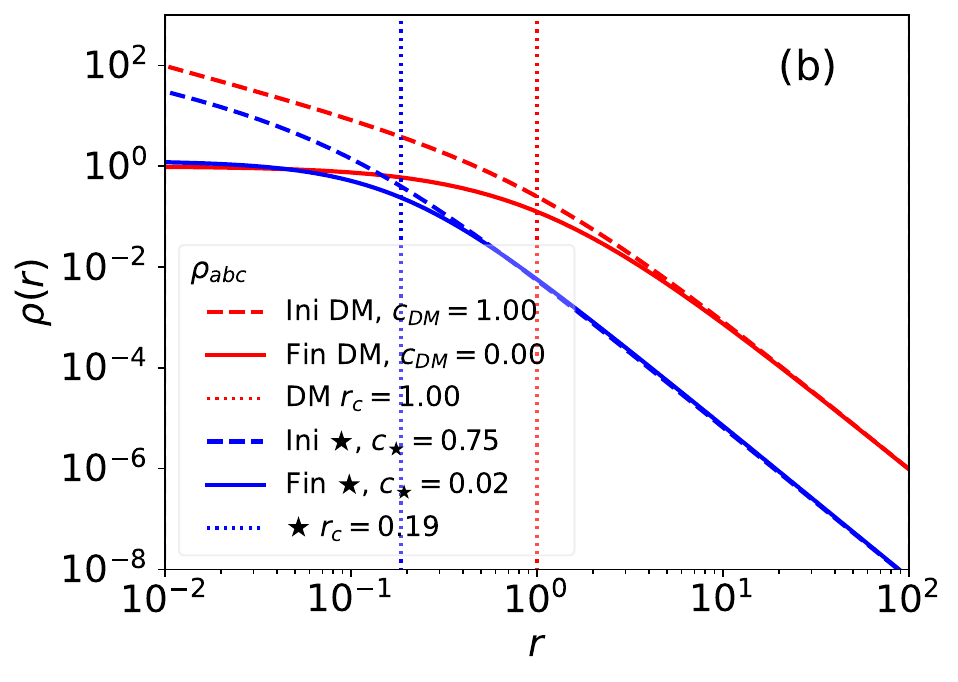}
\includegraphics[width=0.33\linewidth]{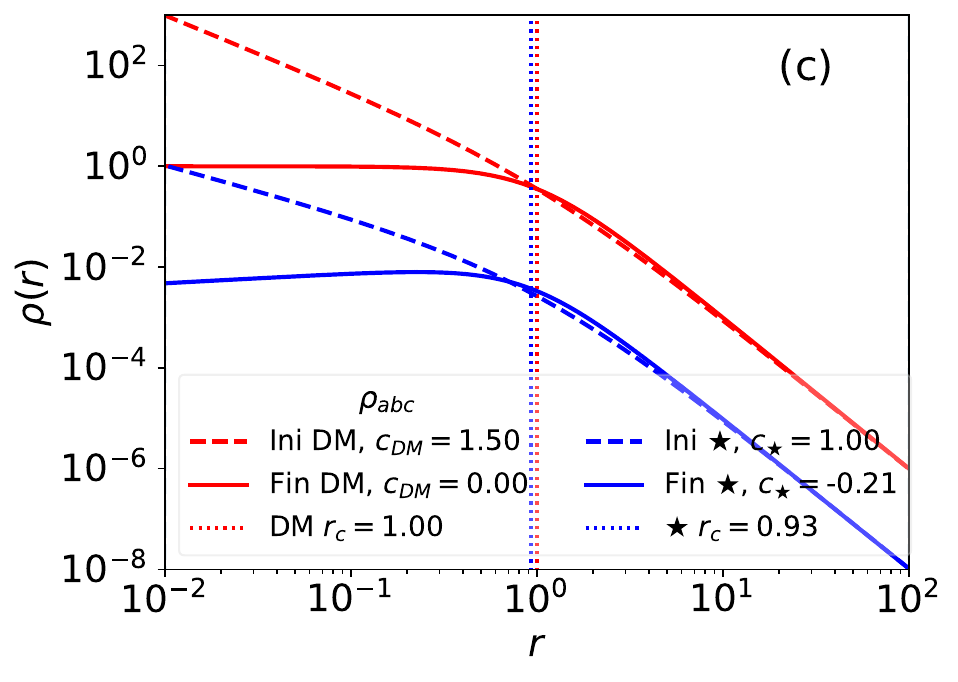}
\includegraphics[width=0.33\linewidth]{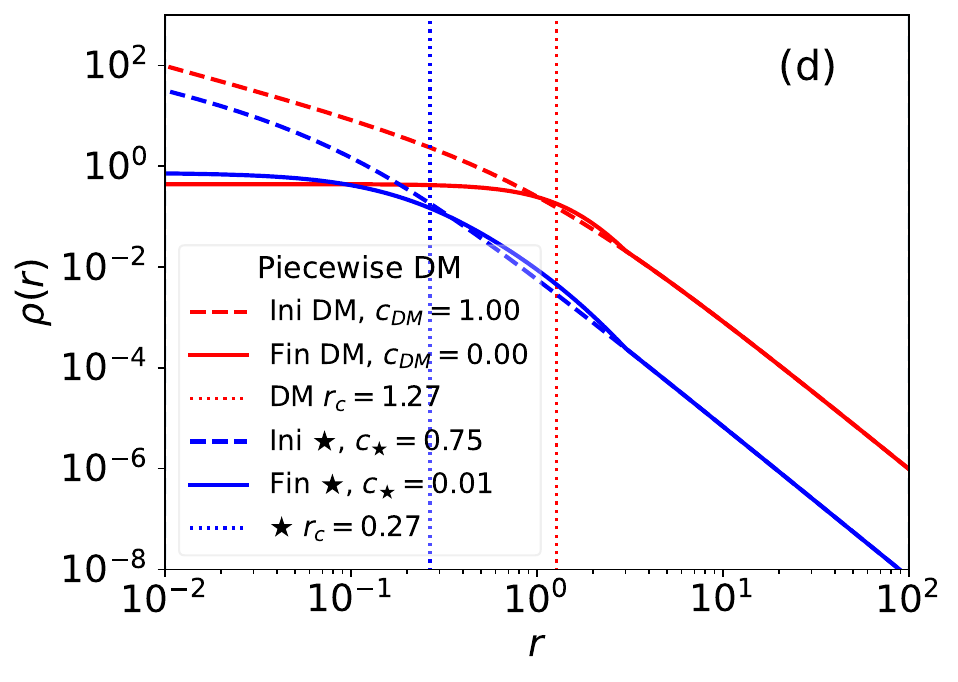}
\includegraphics[width=0.33\linewidth]{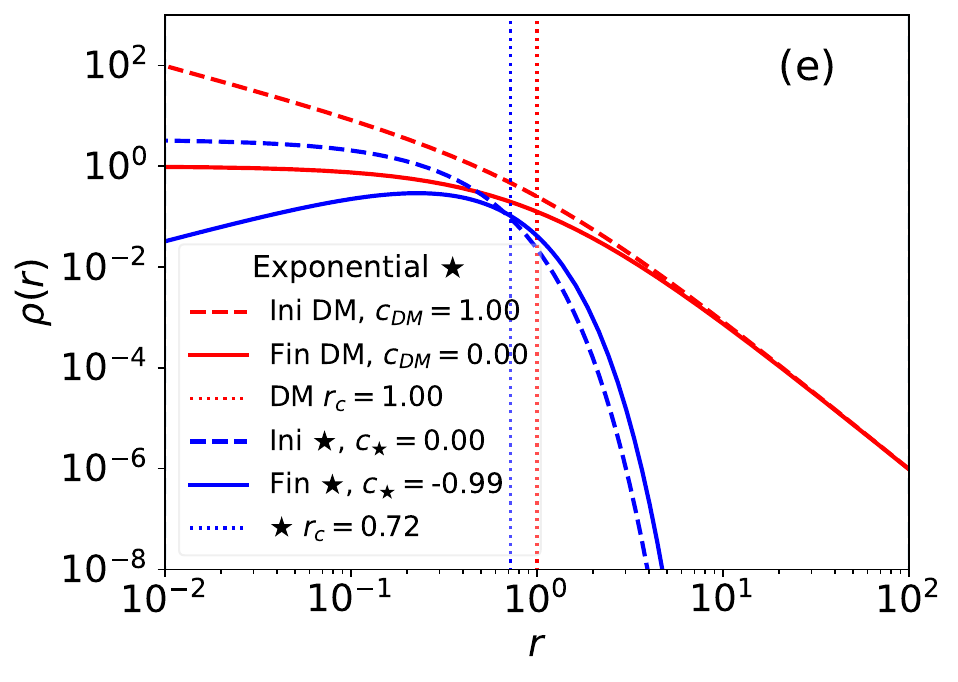}
\includegraphics[width=0.33\linewidth]{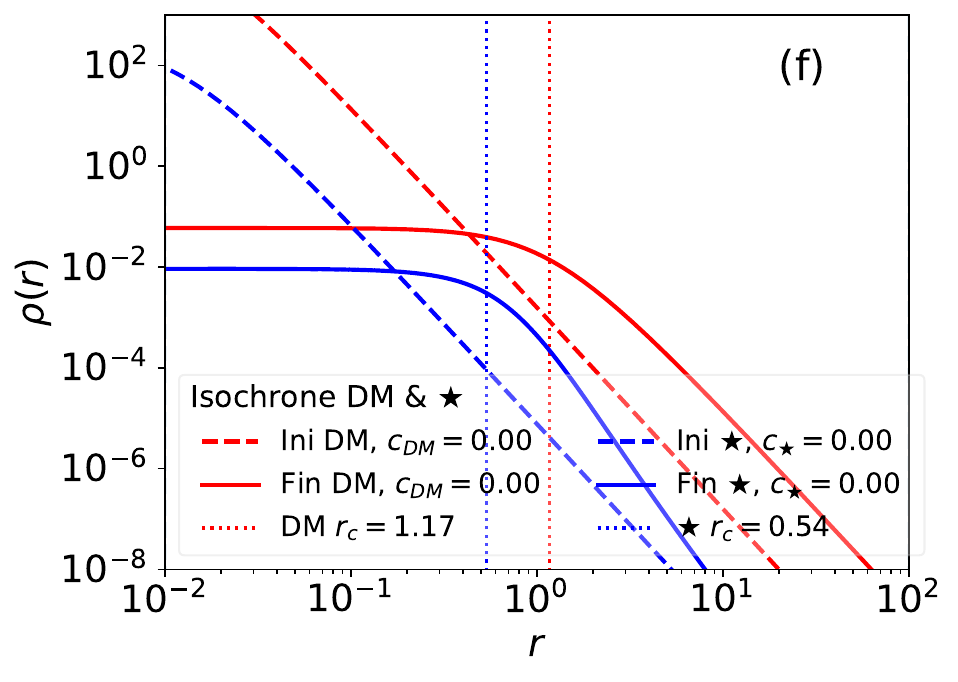}
\caption{
  Examples of the impact of the \dm\ core formation (from the initial red dashed lines to the final red solid lines) on the formation of a stellar core (the blue solid lines) evolving from an originally concentrated stellar distribution (the dashed blue lines). The vertical dotted lines indicate the radius of the formed cores in \dm\ (red) and stars (blue). The initial stars are assumed to follow circular orbits. For details on the types of profiles represented in the different panels, see Sect.~\ref{sec:example_circular}. Arbitrary global scale factors affect both densities and radii.
}
\label{fig:slow_expansion1}
\end{figure*}

We have used Eqs.~(\ref{eq:innermass}), (\ref{eq:scaling}), and (\ref{eq:master}), plugging in specific density profiles, to see the effect on the stars of the \dm\  halo expansion.
Figure~\ref{fig:slow_expansion1}a is used for reference and illustrates the physical process. The DM halo begins as a NFW profile (initial $a,b,c_\dm[t_0] = 1,3,1$) and ends as a cored $abc$ profile where only the inner slope has changed (final $a,b,c_\dm[t] = 1,3,0$). The final total mass has been scaled to be  equal to the initial one. The initial stellar profile is also an $abc$ profile with the inner slope tuned so that at the end it reaches a stellar profile with a perfect core (initial $a,b,c_\star[t_0]=1,3,0.75$; see Eq.~[\ref{eq:constraint1}]). The initial (red dashed line) and final (red solid line)  DM profiles as well as for the initial stellar density profile (blue dashed line), all are assumed to have the same scale radius $r_s$.
To quantify the relative expansion between DM and stars, we use  a core radius $r_c$ defined as the radius where the logarithmic slope of the profile reaches the mean between the maximum and the minimum logarithmic slopes, which are well defined in $abc$ profiles (Eq.~[\ref{eq:abcprofile}]). Explicitly, $r_c$ follows from    
\begin{equation}
  \frac{d\log\rho(r_c)}{d\log r}=\frac{1}{2}\left[\frac{d\log\rho(r)}{d\log r}\Big|_{max}+\frac{d\log\rho(r)}{d\log r}\Big|_{min}\right].
  \label{eq:core_radius_def}
\end{equation}
As Fig.~\ref{fig:slow_expansion1}a shows,  the final stellar distribution (blue solid line; from Eq.~[\ref{eq:master}]) has a core produced by the DM expansion (vertical blue dotted line) similar in size to the final DM profile (vertical red dotted line). For the sake of representation, the stellar mass is assumed to be a hundredth of the \dm\ mass, but this scaling is irrelevant provided the stellar mass is negligible with respect to the \dm\ mass.   The final DM and stellar profiles  show almost exactly the same $r_c$, so that the stellar core is a fair representation of the extent of the \dm\ core.

Figure~\ref{fig:slow_expansion1}b is identical to  Fig.~\ref{fig:slow_expansion1}a except that the initial stellar distribution has a characteristic scale $r_s$ ten times smaller. In this case the stellar core resulting from the expansion is 5 times smaller than the final DM core. The original difference has been reduced but not much. The reason why is analyzed in Appendix~\ref{sec:star_trace_dm}.     

Figure~\ref{fig:slow_expansion1}c is similar to  Fig.~\ref{fig:slow_expansion1}a except for the inner slope of the initial DM profile, $c_\dm(t_0)=1.5$, chosen to be close to the primitive DM halos (see Sect.~\ref{sec:literature}), and for the initial inner slope of the stars, chosen to be $c_\star(t_0)=1$. In addition, the coefficient $a$, that sets the shape of the transition between centers and outskirts in $abc$ profiles, was set to two in the final \dm\ density profile. Note that the inner slope of the final stellar distribution has become slightly positive, in agreement with Eq.~(\ref{eq:other_constraint}).

We also tried three other types of profiles relevant for the work. The first one is shown in Fig.~\ref{fig:slow_expansion1}d and corresponds to the evolution of an initial NFW profile to a piecewise profile having an inner Plummer-Schuster  profile\footnote{It is an $abc$ profile with $a,b,c=2,5,0$.} and the outskirts of the original NFW profile. This kind of evolution is expected in models of \dm\ where the \dm\ particles also interact through forces besides gravity that allow them to collide and thermalize, thus creating an inner core \citep{2021MNRAS.501.4610R,2021MNRAS.504.2832S,2025Galax..13....6S}. In this particular case, the initial distribution of stars is assumed to be more concentrated than the \dm\ and the system evolves to produce a stellar core around five times smaller than the \dm\ core. The radii where the two pieces merge to form the final \dm\ profile is taken to be three times the characteristic scale of the initial NFW \dm\ halo.

Figure~\ref{fig:slow_expansion1}e includes an initial exponential stellar distribution, i.e., $\ln\rho_\star\propto -r$, with the length scale of the exponential much smaller than the scale of the initial \dm\ profile. The expansion of \dm\ creates an inner drop in the stellar mass distribution.  

Finally, Fig.~\ref{fig:slow_expansion1}f shows the effect of \dm\ core formation when the \dm\  is described by a Henon's isochrone potential. It is relevant for the present work because the adiabatic invariants in a Henon's potential are analytical, thus, it allows the analytical study of systems with a wide range of velocity anisotropies. We take advantage of this in  Sect.~\ref{sec:isochrone} to represent stars with isotropic velocities. The functional form of the Henon's isochrone gravitational potential is simple \citep[][]{1959AnAp...22..126H,1959AnAp...22..491H,1960AnAp...23..474H,2008gady.book.....B},
\be
  \label{Isocron}
  \Phi(r) \, = \, -\, \frac{G M_{iso}}{r_{iso}+ A},
\ee
where $r_{iso}$ corresponds to a scale length and $A^2=  r_{iso}^2  +  r^2$. The density profile that generates the isochrone potential is \citep[e.g.,][]{2008gady.book.....B}
 \be
  \label{Isocroden}
  \rho(r) \, =  M_{iso}
  \left[
  \frac{3(r_{iso}+A) A^2 - r^2(r_{iso}+3A)}{4\pi (r_{iso} + A)^3 A^3}
  \right].
\ee
The symbol $M_{iso}$ in Eqs.~(\ref{Isocron}) and (\ref{Isocroden}) stands for the total mass of the density associated with the potential. The density given by Eq.~(\ref{Isocroden}) has a core that vanishes in the limit $r_{iso}\to 0$, and that can be made arbitrarily large increasing $r_{iso}$. Small $r_{iso}$ has been used to represent the initial \dm\ and stellar profiles in Fig.~\ref{fig:slow_expansion1}f; the red and blue dashed lines, respectively.   The generation of the \dm\ core by increasing $r_{iso}$ (the solid red line) induces a stellar core (the solid blue line) which, however, is significantly smaller (compare the vertical blue and red dotted lines).

% 
%%%
\begin{figure*}[ht!] 
\centering
\includegraphics[width=0.8\linewidth]{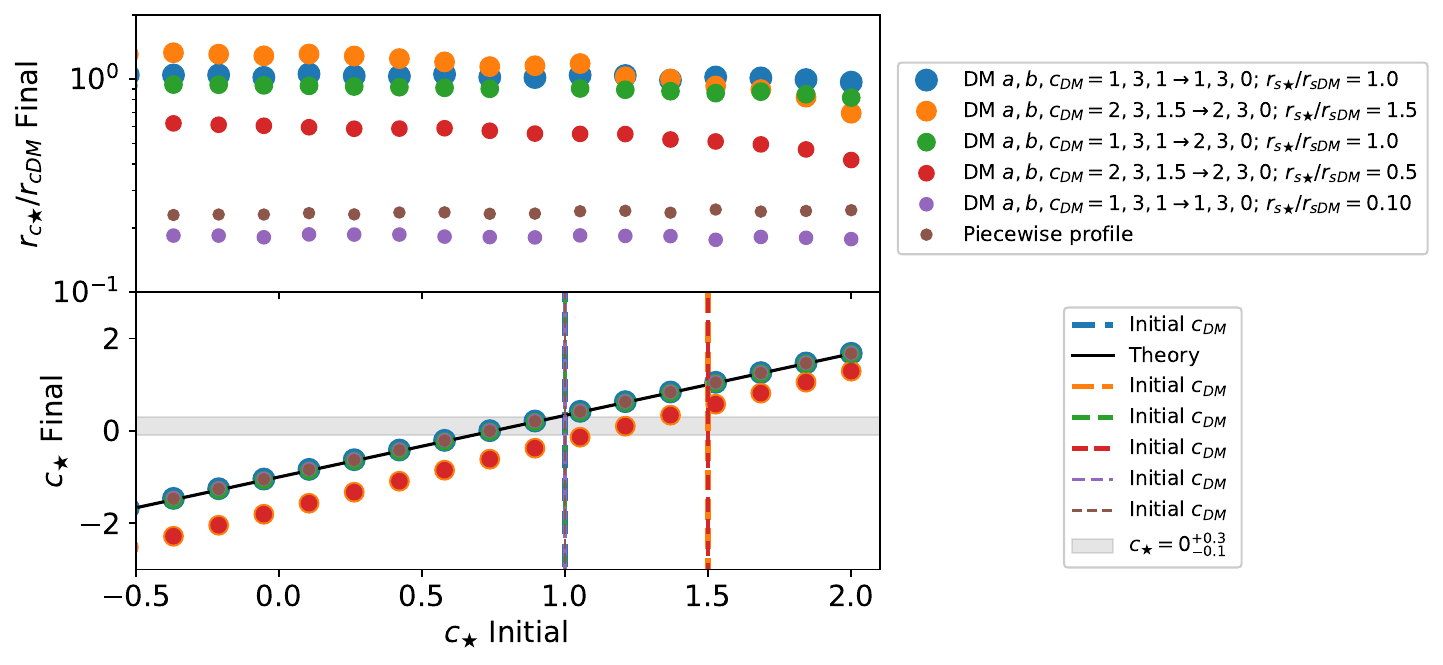}
\caption{Summary of the effect of the \dm\ halo expansion on the stellar distribution when the orbits are circular ($\beta=-\infty$). Two parameters are used to characterize the result: the ratio between the stellar and \dm\ cores (top panel) and the final inner slope of the resulting stellar profile (bottom panel). Both are represented versus the initial inner slope of the stellar distribution. The inset next to the top panel describes the initial and final \dm\ profiles, with $a,b,c_{\dm}$ referring to the parameters defining the $\rho_{abc}$ profiles in Eq.~(\ref{eq:abcprofile}). The black solid line labelled ``Theory'' in the bottom panel represents Eq.~(\ref{eq:slopes}). The vertical dashed lines mark the initial $c_{\dm}$ and follows the same color code as the upper panel. For further details, see Sect.~\ref{sec:example_circular}. The region shaded in gray represents the approximate location of an observed stellar core. 
}
\label{fig:slow_expansion2_plot}
\end{figure*}
The density profiles in Fig.~\ref{fig:slow_expansion1} just provide a glimpse of the effect the \dm\ expansion induces on the stellar distribution. In order to get a more general picture, we carried out a series of simulations varying the properties of the stellar profile for a constant \dm\ density profile variation (same starting and end \dm\ profiles). The results are shown in Fig.~\ref{fig:slow_expansion2_plot}. Each color represents the effect on the stellar distribution of a single pair of initial--final \dm\ profiles. The actual values are given in the inset next to the top panel. Two parameters are used to characterize the expansion; the ratio between the stellar and \dm\ core radii (Fig.~\ref{fig:slow_expansion2_plot}, top panel) and the final inner slope of the resulting stellar profile (Fig.~\ref{fig:slow_expansion2_plot}, bottom panel). Both parameters are represented versus the initial inner slope of the stellar distribution. The main results are:
(1) if the initial \dm\ is described by a NFW profile, then the final stellar density profile develops a core-like feature  (i.e., $c_\star{\rm ~Final}\equiv c_\star[t]\sim 0$) provided the initial stellar distribution has an inner slope in the range  between 0.5 and 1.2 (i.e., $c_\star~{\rm  Initial}\equiv c_\star[t_0]\in [0.5,1.2]$).
(2) The final stellar distribution develops a sort of inner dip or region deprived  of stars when the initial stellar inner slope is well below the 0.5 limit.
(3) The final stellar to \dm\ core radius ratio scales with the initial ratio between the characteristic sizes of \dm\ and stellar distributions, so that the process produces similar \dm\ and stellar cores if the original distributions were of similar in size. In other words, the  expansion rate of the \dm\ is similar to the expansion rate it induces on the stars (see Appendix~\ref{sec:star_trace_dm} for a detailed explanation). Thus, the process of \dm\ expansion can also produce stellar cores smaller and larger than the \dm\ cores depending on the initial conditions.

%
%%%%%%%%%%%%%%%%%%%%%%%
%
\subsection{Initially isotropic stellar orbits in a Henon's isochrone potential}\label{sec:isochrone}

Henon's isochrone potential is defined in Eq.~(\ref{Isocron}). The parameter $r_{iso}$ controls its shape, which can go from the Kepler potential produced by a point mass ($\propto r^{-1}$ when $r_{iso}\to 0$) to the harmonic potential characteristic of a system with constant mass density ($\propto r^2$ when $r_{iso}\to\infty$).  Thus, the density profile that generates the isochrone potential, given in Eq.~(\ref{Isocroden}),  has a plateau when $r \ll r_{iso}$ and becomes a power law ($\propto r^{-4}$) when $r \gg r_{iso}$ \citep[e.g.,][]{2008gady.book.....B}.

Here we consider a dynamical process in which a distribution of stars evolves under the effect of a changing isochrone potential. We assume an initial stellar phase-space distribution $F(r,{\bf v},t_0)$, that is a stationary distribution corresponding to an initial value $r_{iso}(t_0)$. The symbol ${\bf v}$ stands for the velocity vector.  If the parameter $r_{iso}$ changes slowly into a final value $r_{iso}(t)$, the stellar distribution evolves into a corresponding final distribution $F(r,{\bf v},t)$, that is a stationary distribution corresponding to another isochrone potential. Our aim is to determine the form of the final stellar distribution $F(r,{\bf v},t)$, so that the final stellar density can be computed as 
\begin{equation}
  \rho_\star(r,t) = \int \, d^3{\bf v} \, F(r,{\bf v},t).
  \label{eq:density_henon}
\end{equation}
If one  expresses the initial star distribution as a distribution function (\df) of the action variables, then \citep[e.g.,][]{2008gady.book.....B}
\begin{equation}
  F(r,{\bf v},t_0)=\frac{1}{(2\pi)^3}{\mathcal D}(J_r,J_\theta,J_\phi),
  \label{eq:conservation1}
\end{equation}
where ${\mathcal D}$ is the \df\ providing the number of stars having actions in the ranges $(J_r, J_r + dJ_r)$, $(J_{\theta}, J_{\theta} + dJ_{\theta})$, and $(J_{\phi}, J_{\phi} + dJ_{\phi})$. 
The key fact used to determine the final star distribution is that, for a slowly changing potential,  $J_r$, $J_\theta$, and $J_\phi$ are conserved, therefore, ${\mathcal D}(J_r,J_\theta,J_\phi)$ is also conserved, and the final stellar distributions becomes
\begin{equation}
  F(r',{\bf v'},t)=\frac{1}{(2\pi)^3}{\mathcal D}(J_r,J_\theta,J_\phi).
  \label{eq:conservation2}
\end{equation}
We use $r'$ and ${\bf v'}$ rather than $r$ and ${\bf v}$ to evidence that the relation between the action variables and the position and velocity is not the same at $t_0$ and $t$, even though $J_r$, $J_\theta$, and $J_\phi$ are time invariant.
In the case of the Henon's  isochrone potential, the action variables can be written in closed analytical form in terms of the total energy $E$, the total angular momentum $L$, and one of the components of the angular momentum $L_z$ \citep[][]{1991MNRAS.248..494S,2008gady.book.....B},
\be \label{Jr}
J_r \, = \, \frac{GM_{iso}}{\sqrt{-2E}} - \frac{1}{2}
\left[
L + \sqrt{L^2 + 4G M_{iso} r_{iso}},
\right]
\ee
\be \label{Jteta}
J_{\theta} \, = \, L - |L_z|,
\ee
\be \label{Jfi}
J_{\phi} \,= \, L_z.
\ee
\noindent
Assuming isotropic velocities for the initial phase-space stellar \df , then the initial DF depends on $r$ and ${\bf v}$ only through the energy \citep[e.g.,][]{2008gady.book.....B} so that  
\be \label{rvinidist}
F(r,{\bf v},t_0) \, = \, {\mathcal F}[\varepsilon(t_0)],
\ee
\noindent
where $\varepsilon(t_0)$ is the relative energy of a star per unit mass in the initial potential at $t_0$ (i.e., $\varepsilon[t_0] = \, - E$).

The strategy to study the time evolution of the stellar DF consists of translating the initial DF into a distribution for the action variables $J_r$,  $J_\theta$, and $J_\phi$ (Eq.~[\ref{eq:conservation1}]). This distribution preserves its form during the slow change of the parameter $r_{iso}$. Then, at the final time, the action-variables distribution is translated back into
a \df\ of $r'$ and ${\bf v'}$ (Eq.~[\ref{eq:conservation2}]). In practice, because of Eq.~(\ref{rvinidist}),
$\varepsilon (t_0)$  has to be expressed in terms of the action variables in the initial potential, and then  these action variables have to be replaced by the expressions that they adopt in the final potential. We note that the functional dependence of $\varepsilon (t_0)$ on the
action variables explicitly involves  $r_{iso}(t_0)$ for the initial potential and $r_{iso}(t)$ for the final potential. Thus, following this strategy and keeping in mind Eqs.~(\ref{Jr})\,--\,(\ref{Jfi}) with the equivalence
\begin{equation}
\varepsilon(t) = \Psi(r,t)-v^2/2,  
\end{equation}
and
\begin{equation}
  L= r v \sin\eta ,
  \label{eq:def_angular}
\end{equation}
with $\Psi(r,t)=-\Phi(r,t)$, $v=|{\bf v}|$, and $\eta$ the angle between ${\bf r}$ and ${\bf v}$,
$\varepsilon(t_0)$ can be written in terms of the actions expressed in the coordinates of the final potential, namely,
\begin{multline}
\varepsilon(t_0)=\\
\frac{ \Psi(r',t)-v'^2/2}{\left[1+\frac{\sqrt{2\Psi(r',t)-v'^2}}{2GM_t} \Big({\mathcal B}[r',v',\eta,t_0]-{\mathcal B}[r',v',\eta,t]\Big)\right]^2}
\label{eq:long1}
\end{multline}
where,
\begin{equation}
  {\mathcal B}(r,v,\eta,t) = \sqrt{r^2v^2\sin^2\eta + 4G M_t r_{iso}(t)}.
  \label{eq:bbb}
\end{equation}
Then, the final stellar \df\ is just 
\begin{equation}
  F(r',{\bf v'},t)= {\mathcal F}[\varepsilon(t_0)],
  \label{eq:rvinidist2}
\end{equation}
with $\varepsilon(t_0)$ given in Eq.~(\ref{eq:long1}).

We assume a polytropic dependence for ${\mathcal F}$,
\begin{equation}
  {\mathcal F}(\varepsilon)=C\,\left\{\Pi\left[\varepsilon-\Psi(r_{cut},t_0)\right]\right\}^{m-3/2},
  \label{eq:polytrope}
\end{equation}
with $\Pi(x)$ the step function,
\begin{equation}
  \Pi(x)=\begin{cases}
			0, & x \leq 0, \\
            1, & \text{otherwise}.
          \end{cases}
\end{equation}
The assumption on the shape of $\mathcal F(\varepsilon)$ is driven more by mathematical simplicity than by deep physical motivation. However, it is also a common assumption in theoretical work \citep[e.g.,][]{2008gady.book.....B} and naturally appears in the context of self-gravitating systems of maximum entropy \citep{1993PhLA..174..384P}.  (The general case, which does not rely on this assumption, is addressed in  Appendix A.) The constant $-\Psi(r_{cut},t_0)$ in Eq.~(\ref{eq:polytrope}) represents the maximum energy of a star in the initial configuration. Note that all energies are negative and since $\Psi$ is a decreasing function of radius, $r_{cut}$ actually represents the largest radius that a star can reach at $t_0$.  The symbol $m$ stands for the polytropic index and $C$ is a scaling constant assumed to be one in the next equations.

After some lengthy but otherwise straightforward calculation, one can show that the initial density profile arising from Eqs.~(\ref{eq:density_henon}), (\ref{eq:long1}), and (\ref{eq:polytrope}) is given by
\begin{multline}
\rho_\star(r,t_0) =  \\ \pi\, 2^{7/2}\, [\Psi(r,t_0) - \Psi(r_{cut},t_0)]^m \,\int_0^1 (1-x^2)^{m-3/2}x^2dx,
\label{eq:rhoini}
\end{multline}
where we have used that $d^3{\bf v}= 4\pi v^2dv$ since $\varepsilon(t_0)$ just depend on the amplitude of the velocity distribution when $t=t_0$ (Eq.~[\ref{eq:long1}]). 

Similarly, in the case of the final stellar density, Eqs.~(\ref{eq:density_henon}), (\ref{eq:long1}), and (\ref{eq:polytrope}) yield
\begin{equation}
  \rho_\star(r,t) = 
4\pi \, \int_0^1 du 
\int_0^{\sqrt{2\Psi(r,t)}}  {\mathcal G}(r,v,u,t)\,v^2dv,
\label{eq:my_rvfinalpolydit}  
\end{equation}
where we have used  $d^3{\bf v}= 2\pi v^2\sin\eta\, dv d\eta$ and also employ the change of variable $u=\cos\eta$.
The function ${\mathcal G}$ 
 is just ${\mathcal F}[\varepsilon(t_0)]$  except that all variables are brought up explicitly,
  \begin{equation}
 {\mathcal G}(r,v,u,t) = {\mathcal F}[\varepsilon(t_0)],
\label{eq:ggg}
\end{equation}
with ${\mathcal F}$ and $\varepsilon(t_0)$ given in Eqs.~(\ref{eq:polytrope}) and (\ref{eq:long1}), respectively.

The initial velocity distribution was assumed to be isotropic, which allowed us to express the \df\ in terms of the original energy $\varepsilon(t_0)$ (Eq.~[\ref{rvinidist}]). However, the energy of a stellar orbit is not conserved during the expansion ($\varepsilon[t]\not= \varepsilon[t_0]$) which, together with the conservation of the adiabatic invariants leads to final stellar orbits that are not isotropic.  Following the arguments above,  one can also compute the final velocity anisotropy parameter (Eq.~[\ref{eq:defbeta}]).  Since the mean tangential and radial velocities are zero, the tangential and radial velocity dispersions are \citep[e.g.,][]{2008gady.book.....B},
\begin{equation}
\sigma_T^2(r,t)\,\rho_\star(r,t) =\int \, d^3{\bf v}\, (v\sin\eta)^2 \, F(r,{\bf v},t),
\end{equation}
and
\begin{equation}
\sigma_r^2(r,t) \, \rho_\star(r,t) = \int \, d^3{\bf v}\, (v\cos\eta)^2 \, F(r,{\bf v},t),
\end{equation}
which are formally identical to Eq.~(\ref{eq:density_henon}) except for the factors $(v\sin\eta)^2$ and $(v\cos\eta)^2$. Thus, the integration is identical to Eq.~(\ref{eq:my_rvfinalpolydit}), except for these factors, leading to,
\begin{multline}
  \sigma_T^2(r,t) =\frac{4\pi}{\rho_\star(r,t)}  \, \int_0^1du  \int_0^{\sqrt{2\Psi(r,t)}} dv\,\times \\
  (1-u^2)\,v^4\, {\mathcal G}(r,v,u,t),
  \label{eq:sigma_t}
\end{multline}
and
\begin{multline}
  \sigma_r^2(r,t) =\frac{4\pi}{\rho_\star(r,t)}  \, \int_0^1du  \int_0^{\sqrt{2\Psi(r,t)}} dv\,\times \\
  u^2\,v^4\, {\mathcal G}(r,v,u,t).
  \label{eq:sigma_r}
\end{multline}

The above equations are used in the next section 
%Sect.~\ref{sec:example_isotropic}
to evaluate stellar density profiles and anisotropy parameters characteristic of stellar system with isotropic velocity distributions.

%%%
% the asterisk removes the numbering, as required by Pascale Monier, A&A Editorial assistant 
\subsubsection*{Examples of systems with initial isotropic orbits}\label{sec:example_isotropic}

\begin{figure*}[ht!] 
\centering
\includegraphics[width=0.49\linewidth]{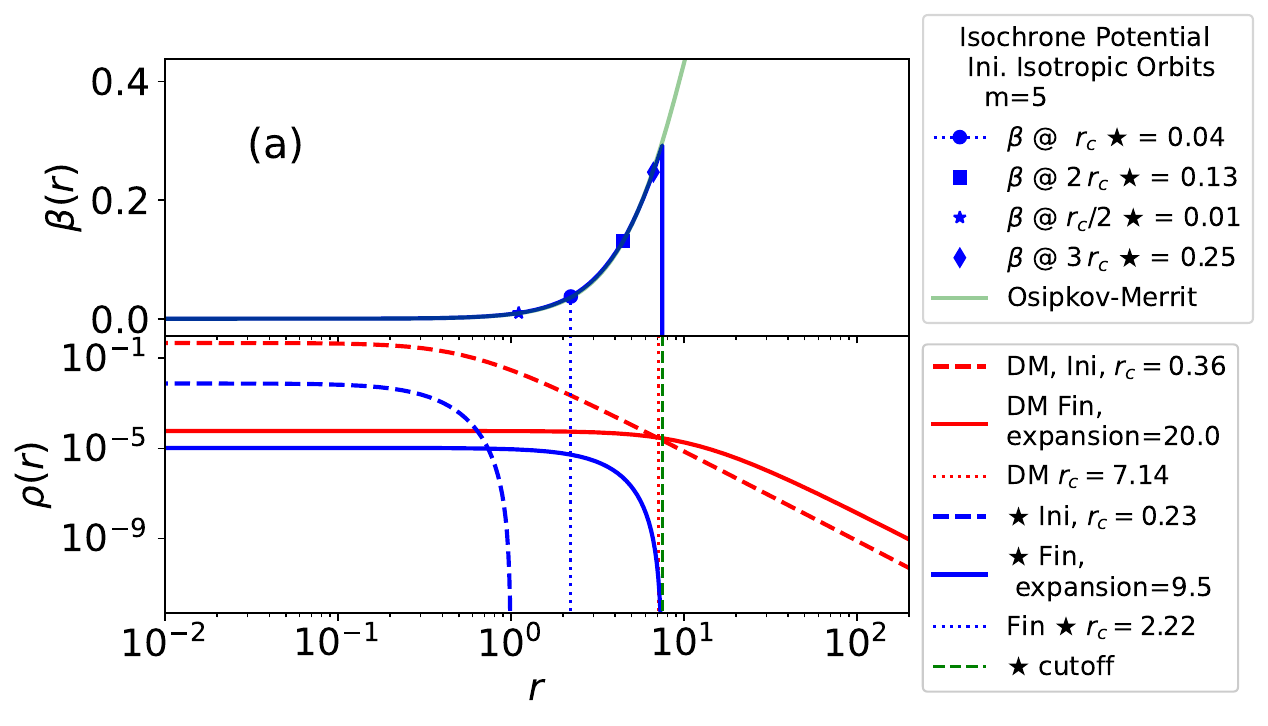}
\includegraphics[width=0.49\linewidth]{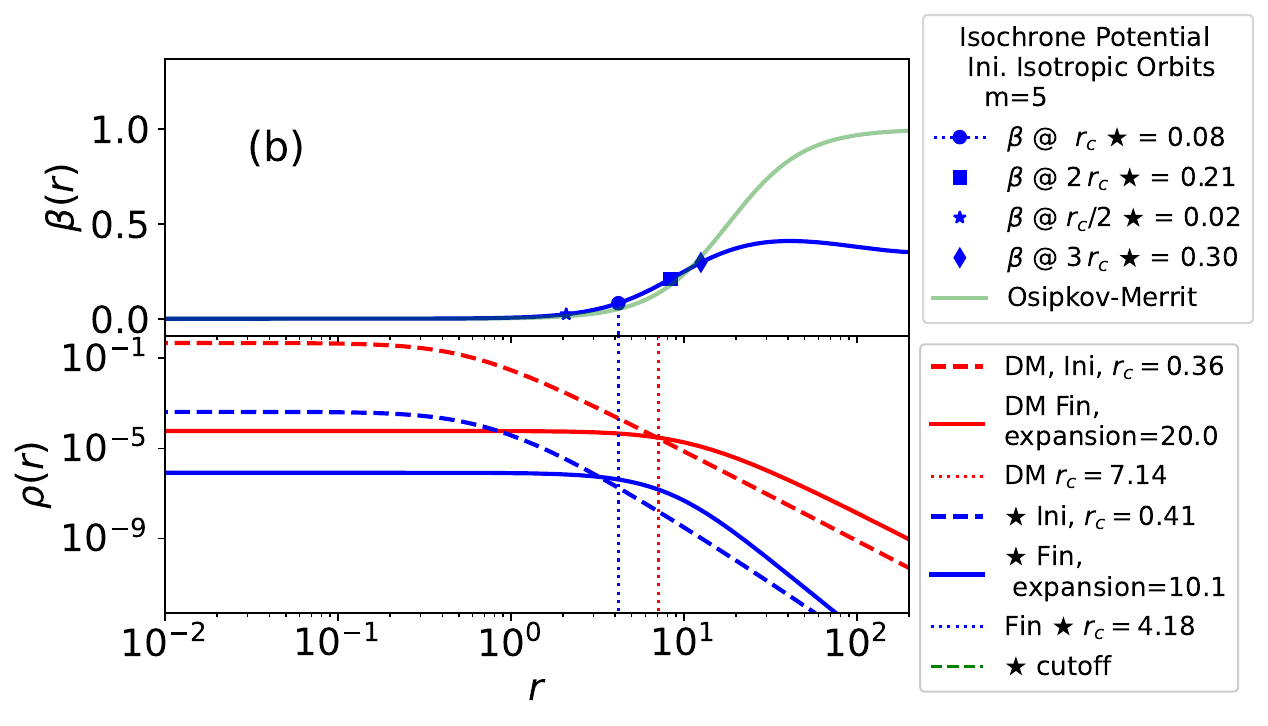}
\includegraphics[width=0.49\linewidth]{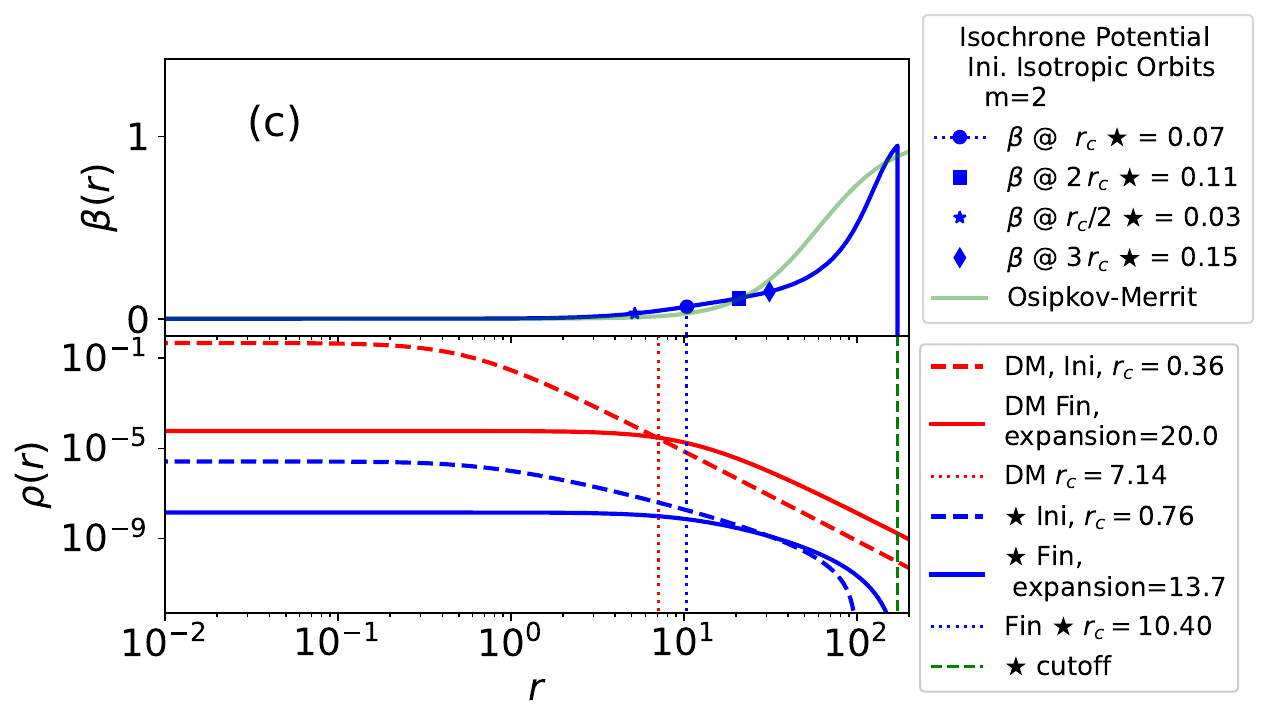}
\includegraphics[width=0.49\linewidth]{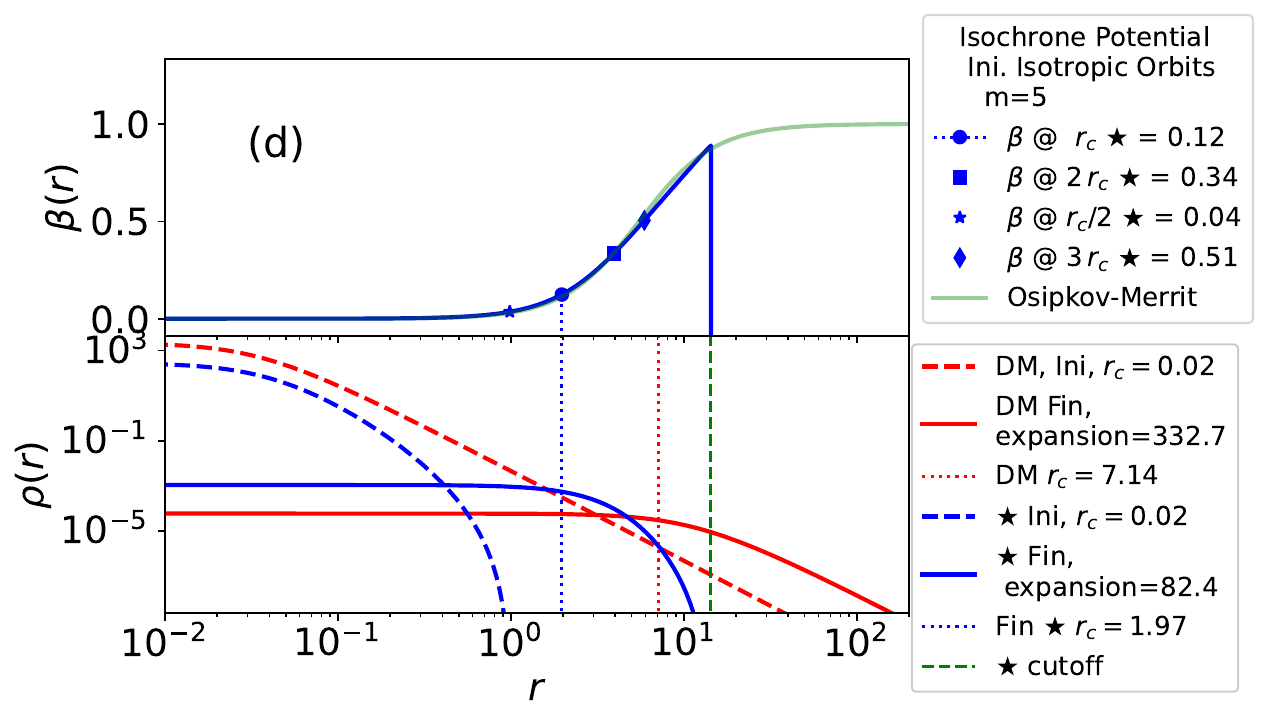}
\caption{Simulations corresponding to an initially isotropic stellar distribution being hosted in a \dm\ halo set by Henon's isochrone potential.  Each figure is split into two panels with the anisotropy parameter on top and the density profiles and core radii at the bottom.  The color and line-type code for the densities (bottom panels) is the same as that used in Fig.~\ref{fig:slow_expansion1} and \ref{fig:slow_expansion5}. The inset in the top panel includes the actual value of $\beta$ at the core radius $r_c$ of the final stellar distribution, as well as $r_c/2$, $2\,r_c$, and $3\,r_c$. 
(a)~The \dm\ expands by a factor of 20. The polytropic index $m$ is set to 5, and the cutoff radius of the initial stellar distribution to 1. The cutoff of the final stellar distribution is marked with a vertical green dashed line.  
(b)~Same as (a) except for the initial cutoff, set to 1000.
(c)~Same as (a) except for the initial cutoff, set to 100, and the polytropic index, set to 2.
(d)~Same as (a) except for the \dm\ expansion factor, which this time is about 350 ($r_{iso}$ goes from 0.03 to 10). For reference, the four panels include an Osipkov-Merrit (OM) anisotropy distribution (the green solid lines), where $\beta(r)=r^2/(r^2+r_{\rm OM}^2)$.  In general, OM does a fair job when the characteristic radius,  $r_{\rm OM}$, is tuned to a value much larger than the DM core radius. For further details on the types of line in the density sub-panels, see the caption of Fig.~\ref{fig:slow_expansion1}.
}
\label{fig:slow_expansion6}
\end{figure*}

After setting the parameters that define the initial and the final \dm\ profiles, Eqs.~(\ref{eq:defbeta}), (\ref{eq:rhoini}), (\ref{eq:my_rvfinalpolydit}), (\ref{eq:sigma_t}), and (\ref{eq:sigma_r}) are used to evaluate the final stellar density profiles and the anisotropy parameters shown in Fig.~\ref{fig:slow_expansion6}. The single and double integrals in the equations were evaluated numerically at each radius $r$ using the routine {\tt simpson} from the Python package {\tt SciPy} \citep{2020SciPy-NMeth}.

The top panel within each sub-figure of  Fig.~\ref{fig:slow_expansion6} shows the anisotropy parameter of the final stellar distribution, whereas the bottom panel contains the relevant  densities and core radii. By construction, the \dm\ density creating the potential has a core whose size is set by $r_{iso}$ and can be made small to represent initial conditions and large to represent the final ones  (see the dashed and solid red lines in the bottom panels of Fig.~\ref{fig:slow_expansion6}). The \dm\ cores are inherited by both the initial and the final stellar distributions (blue dashed lines and blue solid lines, respectively). We parameterize the expansion rate, for \dm\ and for stars, as the ratio between the initial and final core radii defined in Eq.~(\ref{eq:core_radius_def}). 
The typical expansion rates of the stars are large but smaller than the expansion rate of the \dm\ distribution that causes the stellar expansion. Stars expand about half the expansion of the \dm , as it is registered in the inset of the different panels. The final stellar velocity anisotropy  happens to be isotropic in the stellar core ($\beta\simeq 0$ for $r\lesssim r_c$) and becomes radially biased in the outskirts of the stellar profile ($\beta\gtrsim 0$ for $r\gtrsim r_c$). Such radial anisotropy is never large within potentially-observable radii. The shape resembles the  Osipkov-Merrit anisotropy used in analytic studies, where  $\beta(r)=r^2/(r^2+r_{\rm OM}^2)$ with $r_{\rm OM}$ a characteristic length scale; see the green curve in the top panels of Fig.~\ref{fig:slow_expansion6}.

\begin{figure}[ht!] 
\centering
\hspace*{-0.3cm}\includegraphics[width=1.05\linewidth]{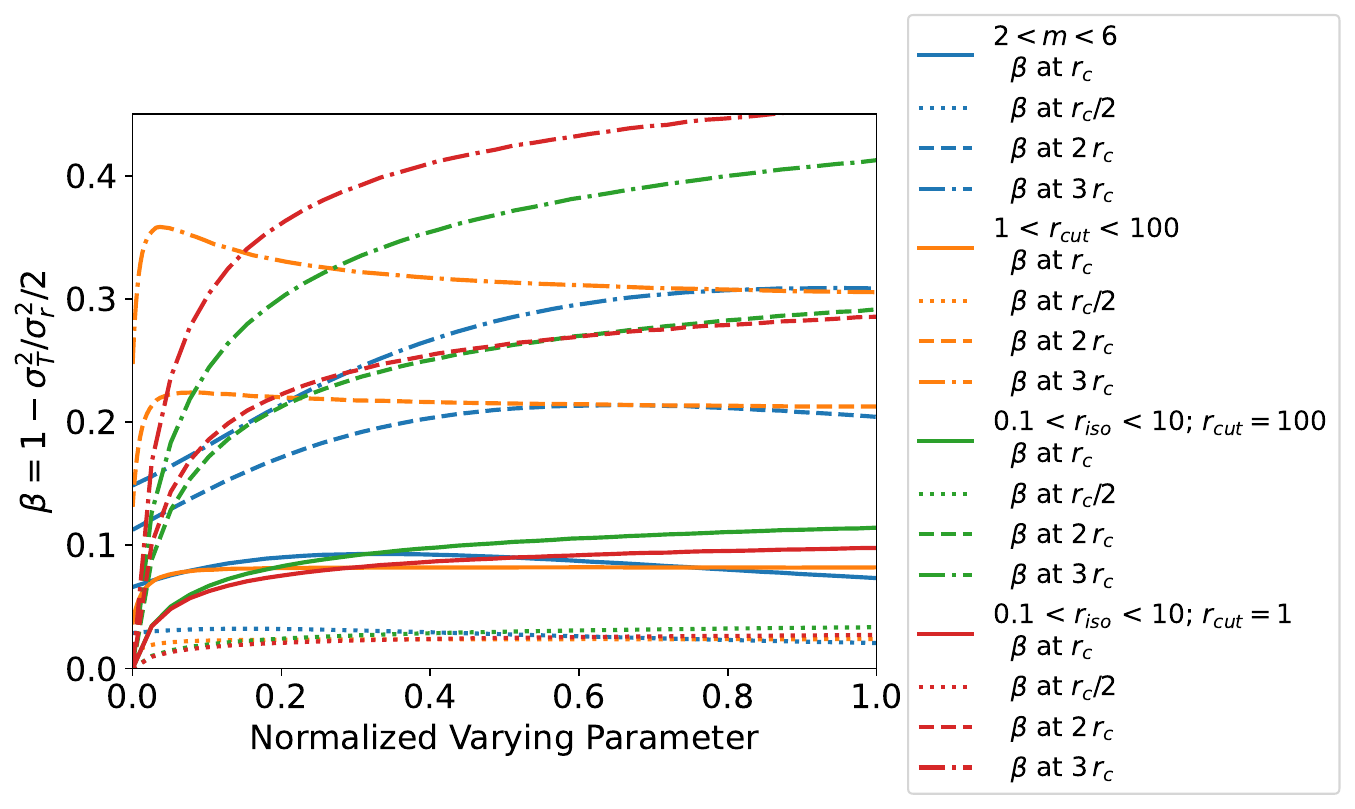}
\caption{Variation of the anisotropy parameter $\beta$ at selected radii of the final stellar profile:  at the core radius $r_c$ (the solid lines), within the core radius at $r_c/2$ (the dotted lines) and outside the core at $2\,r_c$ and $3\,r_c$ radii (the dashed and dotted-dashed lines, respectively). Different colors represent different varying parameter as indicated in the inset. The absolute variation, given in the inset, has been normalized so that if $p$ is the varying parameter then the abscissa represents $(p-\min p)/(\max p-\min p)$.
  The gravitational potential set by \dm\ is an Henon's isochrone with an initial stellar velocity distribution that is isotropic (Sect.~\ref{sec:isochrone}).
}
\label{fig:slow_expansion7}
\end{figure}

In order to quantify more precisely the velocity anisotropy  produced by the DM expansion, we carried out a series of simulations where $\beta$ is evaluated at different radii of the resulting stellar profile: at the core radius $r_c$, within the core radius at $r_c/2$, and outside the core at $2\,r_c$ and $3\,r_c$ radii. The results are condensed in Fig.~\ref{fig:slow_expansion7}, where  $\beta$ is represented versus the free parameter that is varied in the simulation. We vary the polytropic index ($2\le m \le 6$; the blue lines), the stellar cutoff radii ($r_{cut}$, the orange lines), and the final \dm\ core radius ($r_{iso}$, with $r_{cut}$ fixed to 1 and 100; the read and green lines, respectively). The simulations show that within the core $\beta\simeq 0$, with a typical value at $r_c$ around 0.07. Outside a core, $\beta$ is positive and increases to typical values between 0.2 and 0.5. 

\begin{figure}[ht!] 
\centering
\hspace*{-0.3cm}\includegraphics[width=1.05\linewidth]{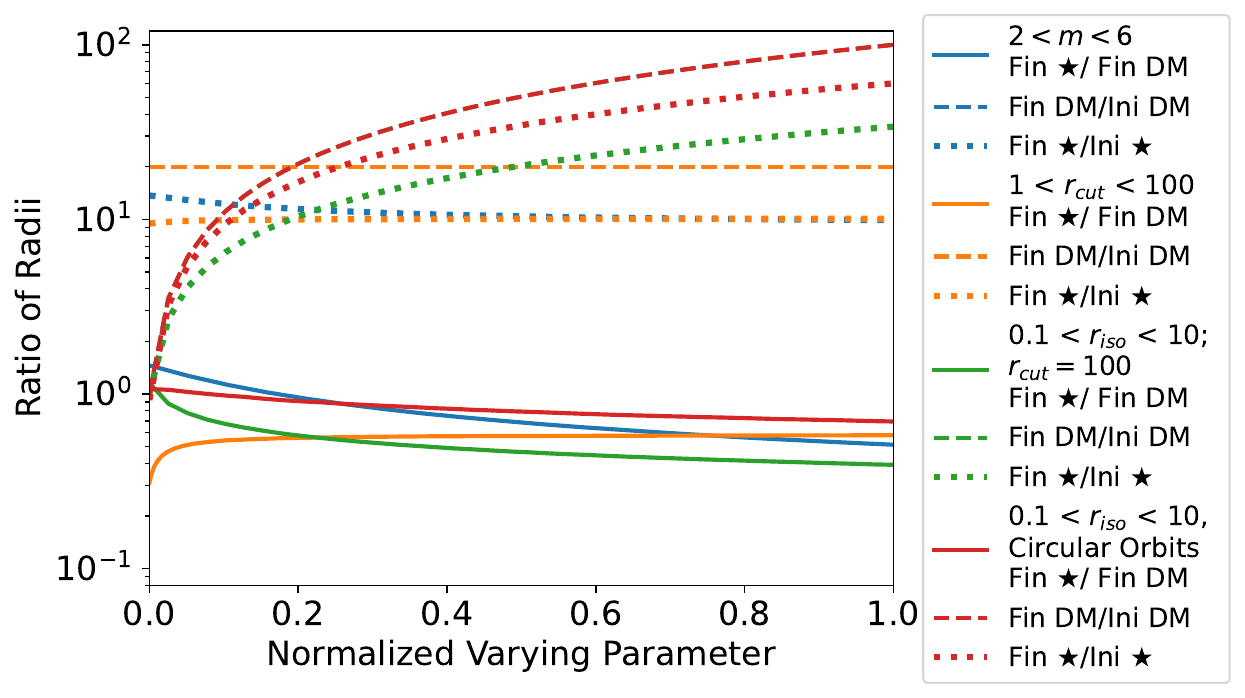}
\caption{
 Ratio between radii characterizing the effect of the expansion of the \dm\ halo on an initially isotropic stellar velocity distribution. It shows final core radius ratio $\star$\,$r_c$/\dm\,$r_c$ (the solid lines), the final over initial core ratio of \dm\, (the driving expansion rate, given by the dashed lines), and  the final over initial core ratio of the stars (the resulting stellar expansion rate, given by the dotted lines). They are presented versus the variation of the polytropic index (the blue lines), the cutoff radius of the initial stellar distribution (the orange lines), the initial core radius of \dm\ (the green lines), and  the initial core radius of \dm\ but assuming circular orbits (Sect.~\ref{sec:spherical}; the red lines). The gravitational potential  is an Henon's isochrone, and the varying parameter has been normalized as explained in the caption of Fig.~\ref{fig:slow_expansion7}.
}
\label{fig:slow_expansion4_plot}
\end{figure}

As far as the stellar density is concerned, the general behavior is summarized in Fig.~\ref{fig:slow_expansion4_plot}. It shows the final core radius ratio $\star$\,$r_c$/\dm\,$r_c$ (the solid lines), the final over initial core ratio of \dm\, (the dashed lines), and  the final over initial core ratio of the stars (the dotted lines). They are presented versus the variation of the polytropic index (the blue lines), the cutoff radius of the initial stellar distribution (the orange lines), the initial core radius of \dm\ (the green lines), and  the initial core radius of \dm\ but assuming circular orbits as worked out in Sect.~\ref{sec:spherical} (the red lines). The main result is that the stellar expansion is smaller but of the same magnitude as that \dm\ expansion driving it, and that the resulting stellar core is smaller but of similar size as the final \dm\ core.     
We argue in Appendix~\ref{sec:star_trace_dm} that the expansion rate of the stars scales with the expansion rate of \dm\ as a power law, with an exponent close to one (around 0.85; see Eq.~[\ref{eq:expanssion_iso}]). 

%
%%%%%%%%%%%%%%%%%%%%%%%%%%%%%%%%%%%%%%%%
\subsection{Radial stellar orbits in arbitrary spherical potentials}\label{sec:radial}
\begin{figure}[ht!] 
\centering
\includegraphics[width=0.8\linewidth]{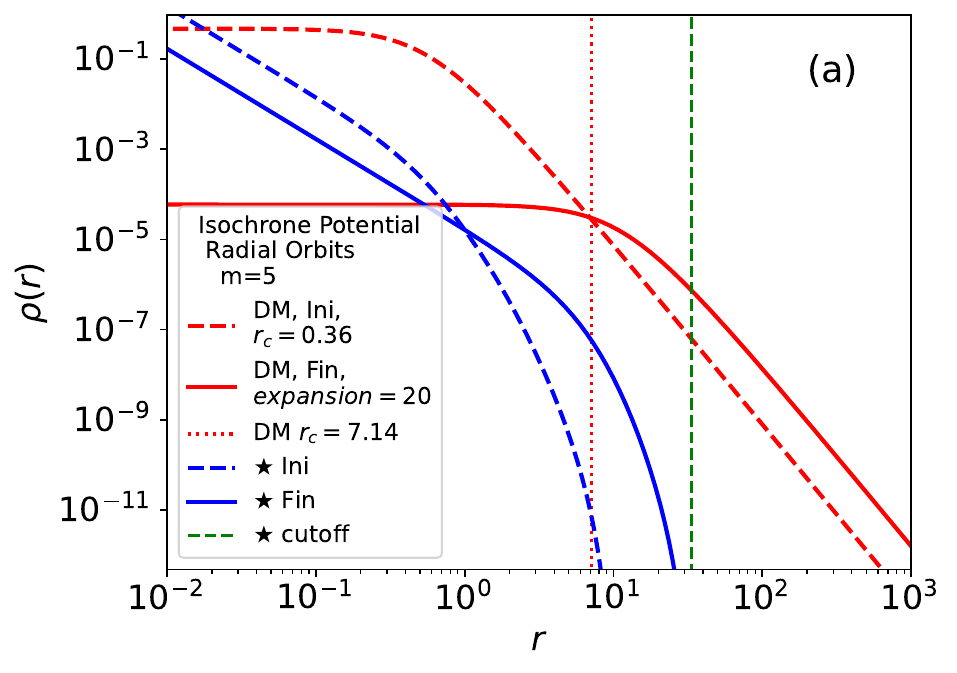}
\includegraphics[width=0.8\linewidth]{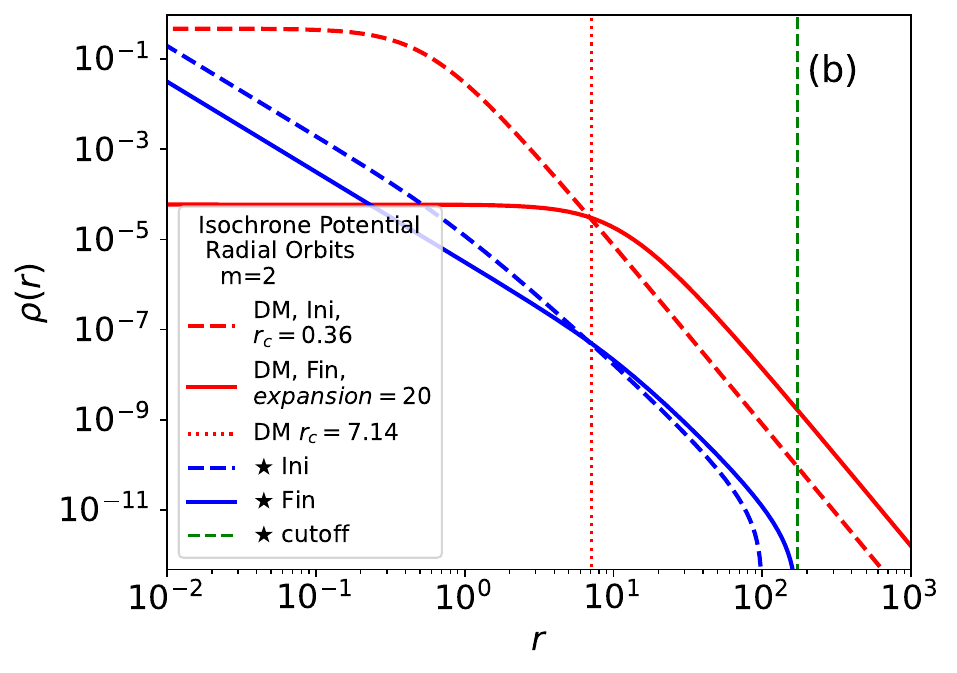}
\caption{Similar to Figs.~\ref{fig:slow_expansion1} and \ref{fig:slow_expansion6}, except that the orbits are radial thus preventing the formation of a stellar core (see the blue solid lines, which behave as $r^{-2}$ at small radii). The potential was assumed to be of the type Henon's isochrone. 
(a)~The \dm\ expands a factor of 20, from $r_{iso}=0.5$ to $r_{iso}=10$. The polytropic index $m$ is set to 5, and the cutoff radius of the initial stellar distribution to 10. The cutoff of the final stellar distribution is marked with a vertical green dashed line.  
(b)~Same as (a) except for the initial cutoff, set to 100, and the polytropic index, set to 2. 
For further details, see the caption of Fig.~\ref{fig:slow_expansion1}.
}
\label{fig:slow_expansion5}
\end{figure}
In stellar systems with radial orbits, the angular momentum is zero ($L$ in Eq.~[\ref{eq:def_angular}]) since the radial vector and the velocity are defined to be parallel. Because the angular momentum is conserved during the assumed adiabatic expansion of the \dm\ halo, the final stellar distribution also has radial orbits. Since $L=0$, the dependence of both the initial and the final DF on $L$ is through a Dirac $\delta_+(L^2)$ function \citep[][problem 4.9]{1984pgs1.book.....F,2008gady.book.....B},  explicitly,
\begin{equation}
 F(r,{\bf v}) =  \delta_+(L^2)\, {\mathcal F}({\epsilon})
  =  \delta_+(L^2) {\mathcal F}[\Psi(r) - v^2/2].
\label{eq:dirac2}
\end{equation}
 It follows directly from the above DF, by integrating over the velocities, that
 the spatial density is
\begin{equation}
  \rho_\star(r) = \frac{2\pi}{r^2} \int_0^{\sqrt{2\Psi(r)}} dv\, {\mathcal F}[\Psi(r) - v^2/2].
\end{equation}
We then see that stellar systems with purely radial orbits do not admit stellar cores,
since the density scales as $\rho(r) \propto r^{-2}$. These stellar systems are, however,
rather artificial. Small perturbations lead to orbits that are not strictly
radial that blur the central cusp.

For the purpose of illustration, we worked out the initial and final densities in a Henon's potential (Eq.~[\ref{Isocron}]), that allows for radial stellar orbits. In this case, Eq.~(\ref{eq:dirac2}) becomes,
\begin{equation}
F(r,{\bf v},t) = \delta_+(L^2)\,{\mathcal F}[\varepsilon(t_0)],
\label{eq:dirac}
\end{equation}
where we also assume a dependence of the DF on $J_r$ identical to that used in Sect.~\ref{sec:isochrone}. 
Inserting the DF, Eq.~(\ref{eq:dirac}), into Eq.~(\ref{eq:density_henon}),  one can show that the initial density is
\begin{multline}
  \rho_\star(r,t_0) = \frac{\pi\, 2^{3/2}}{r^2}\times\\ \left[\Psi(r,t_0)-\Psi(r_{cut},t_0)\right]^{m-1}\,\int^1_0(1-x^2)^{m-3/2}dx,
  \label{eq:radial2}
\end{multline}
whereas the final density is %\comment{revise}
\be
\rho_\star(r,t)  =   \frac{2\pi}{r^2} \,\int_0^{\sqrt{2\Psi(r,t)}}  {\mathcal G}(r,v,1,t)\,dv.
\label{eq:radial1}
\ee
Setting $u=1$ in ${\mathcal G}$   (see the definition in Eq.~[\ref{eq:ggg}]) corresponds to $\sin\eta =0$ and, consequently, to $L=0$ (Eq.~[\ref{eq:def_angular}]).

Examples based on solving numerically Eqs.~(\ref{eq:radial2}) and (\ref{eq:radial1}) are shown in Fig.~\ref{fig:slow_expansion5}. Contrarily to what happens with circular orbits (Fig.~\ref{fig:slow_expansion1}) and isotropic orbits (Fig.~\ref{fig:slow_expansion6}), the inner slope of the final stellar distribution never forms a core (the blue solid lines) despite the fact that the final \dm\ profile has it (the red solid lines).

%
%%%%%%%%%%
%_____________________________________________________________
%                                             Table with notes
%-------------------------------------------------------------
% Single note
%-------------------------------------------------------------
\begin{table*}[h!]
\caption{Constraints on the expected initial and final conditions \label{tab:literature}}
\centering
\begin{tabular}{lcccl}
\hline\hline
Parameter & Symbol&Range$^1$ & Definition & Comment\\
\hline
Initial \dm\ inner slope & $c_\dm(t_0)$& 1.5\,--\,2& $c$ in Eq.~(\ref{eq:abcprofile})& Quite robust\\
Initial \dm\ radius &$r_s$&60\,--\,260\,pc&Eq.~(\ref{eq:abcprofile})&$10^6$--$10^8 M_\odot$ halo mass\\
Initial stellar inner slope&$c_\star(t_0)$& $0$\,--\,$c_\dm(t_0)$& $c$ in Eq.~(\ref{eq:abcprofile})&Very uncertain\\
Initial stellar radius&$r_{iso}, r_s$&3\,--\,150\,pc&Eqs.~(\ref{Isocroden}) \& (\ref{eq:abcprofile})  & Uncertain\\
Initial stars to \dm\ radii ratio&---&0.1\,--\,1.5&---& $>0.4$, Appendix~\ref{app:blablabla} \\
Final central velocity anisotropy&$\beta(0)$& $\sim$ 0 & Eq.~(\ref{eq:defbeta})& Observations \& simulations \\
Final velocity anisotropy &$\beta(r)$& 0\,--\,$\pm 1$& Eq.~(\ref{eq:defbeta})& Observations \\
Final velocity anisotropy &$\beta(r)$& 0, 0\,--\,$1$& Eq.~(\ref{eq:defbeta})& Simulations\\
\hline
\end{tabular}
\tablefoot{$^1$Range of values from the literature reviewed in Sect.~\ref{sec:literature}.}
\end{table*}

\section{Initial conditions to be expected}\label{sec:literature}

As it was described in Sect.~\ref{sec:models}, under quite general circumstances, the expansion of the \dm\ halo forces the expansion of the initial stellar distribution that forms a core-like feature. However, the properties of the resulting stellar cores depend on the initial conditions of the distribution of \dm\ and stars, including their relative size  and the anisotropy of the stellar orbits.
Thus, we address the question of which initial conditions can be expected, based on the existing literature. The result is discussed next and also summarized in Table~\ref{tab:literature}.

\noindent{\em Constraints on the initial \dm\ distribution:} According to the existing C\dm\ numerical simulations, the first \dm\ haloes have a power-law mass density distribution with the inner slope, $c_\dm(t_0)$,  being around 1.5 or 2 \citep{2004ApJ...607..125T,2005Natur.433..389D,2017ApJ...846...30Y,2021A&A...647A..66C,2023MNRAS.518.3509D,2023JCAP...10..008D}, which is significantly larger than the NFW slope ($c=1$ in Eq.~[\ref{eq:abcprofile}]). These high-redshift halos are also small, with typical concentrations ($r_s/r_{vir}$) around $\sim 5$ independently of the \dm\ halo mass \citep{2013JCAP...04..009A,2014ApJ...788...27I,2014MNRAS.441.3359D,2015MNRAS.452.1217C,2025MNRAS.536..728S}. The symbol  $r_{vir}$ stands for the  virial radius which  depends on the total halo mass and the cosmological parameters so that, for the current cosmology and for halo mass from $10^6$ to $10^8 M_\odot$, $0.3 < r_{vir}< 1.3\,{\rm kpc}$ at redshift ten.\footnote{For the current cosmology, with $\Omega_m=0.3$ and $\Omega_\Lambda=0.7$, $r_{vir}\simeq 9.6\,{\rm kpc}\,[M_\dm(<r_{vir})/10^8\,M_\odot]^{1/3}\,[0.7+0.3\,(1+z)^3]^{-1/3}$, with $z$ the redshift and $M_\dm(<r_{vir})$ the \dm\ halo mass within the virial radius \citep[e.g., see ][]{2025A&A...694A..17Q}. %\comment{revise. Eqs in wiki under 'virial radius'}
}
Thus, the typical initial scale radii, $r_s$, span from 60 to 260\,pc.  These estimates are included in Table~\ref{tab:literature}.

These conditions are derived from C\dm\ simulations and, in principle, may not directly apply to other \dm\  models where the expansion is significant (Sect.~\ref{sec:intro}). However, the expansion we propose unfolds over long timescales, suggesting that the physical mechanism behind it is unlikely to play a major role until later epochs.  At least, this is our working hypothesis so that the C\dm\ initial conditions are assumed to hold for other types of \dm .

\noindent{\em Initial stellar distribution:}
There is no clear consensus in the literature regarding the stellar distribution that emerges as \dm\ halos first form stars.
In relatively massive objects going through major star formation episodes,  the stellar surface density seems to be a power law at all times, with an exponent as negative as -2 \citep{2023MNRAS.522.4515L}. It remains to be determined how representative these starbursts are of star formation in smaller \dm\  halos.
As for the size, several recent numerical simulation impose stellar distributions with half-mass radius similar to the size of the dark matter distribution mentioned in the previous paragraph. For example, \citet{2022MNRAS.515..302R} assume the stellar half-mass radius to \dm\ virial radius of 0.15. With the typical concentrations of high redshift \dm\ halos ($\sim 5$), this ratio is equivalent to a ratio between stars and \dm\ radii of   $\sim 0.75$, signaling their similarity.  \citet{2022MNRAS.515..302R} consider uncertainties exploring ratios in the range between 0.15 and 1.5.
\citet{2025A&A...694A..17Q} use initial (redshift 7) stellar distributions between 15 and 150\,pc when the DM halo mass spans from $10^{7.5}$\,--\,$10^8\,M_\odot$. Thus, the assumed initial stellar distribution is smaller but of the order of the $r_s$ defining the \dm\ distribution.
In short, the stellar distribution is likely a power law with negative exponent and a spatial extent smaller but not very different from the extent of the dark matter halos. These constraints agree with our argument in Appendix~\ref{app:blablabla} that the half-mass radius of the initial stellar distribution has to be larger than 0.4 times the half-mass radius of the initial \dm\ halo.  

Another independent physical constraint on the original stellar distribution comes from the newborn star forming regions observed locally. We focus on large H\,{\sc ii} regions because they likely reflect the physical conditions characteristic of the early Universe, with abundant high-density gas ready to form stars. Thus, local analogs to high redshift star-formation could be the so-called super star clusters (SSC) observed to happen in, e.g., major mergers of gas rich galaxies \citep{1995AJ....109..960W}. These SSCs are young ($100$\,Myr), have stellar mass around $10^5\, M_\odot$ \citep{2005ApJ...619..270M}, and cored light distributions with half-light radii from a 3 to 10 pc \citep{2020MNRAS.492..993C}. Compared to the sizes of the initial \dm\ distributions described above, these stellar radii represent a fraction going from approximately 0.1 to 1.5 times the \dm\ radii.  These estimates are included in Table~\ref{tab:literature}.

\noindent{\em Stellar velocity anisotropy:}
One of the conclusions reached in Sect.~\ref{sec:models} is that an initial tangential or radial velocity anisotropy remains tangential or radial after the expansion. However, initially isotropic velocity distributions turn into isotropic in the formed core but radially biased beyond the core radius (Sect.~\ref{sec:example_isotropic} and Fig.~\ref{fig:slow_expansion6}). Thus, according to the \dm\ expansion mechanism, the final anisotropy of the stellar velocity keeps memory of the initial conditions. Therefore, the question arises: what is the observed anisotropy? The velocity anisotropy observed in dwarf late type galaxies is quite uncertain and rely on modeling the velocity and mass distribution in dSph and UFDs. However, the consensus is that galaxies with cores have isotropic stellar velocities at the center \citep[$\beta\sim 0$;][]{2009MNRAS.394L.102L,2002MNRAS.333..697L,2016MNRAS.463.1117Z,2018MNRAS.480..927P,2022ApJ...939..118K,2022A&A...659A.119K}. In order to be consistent with the proposed mechanism, the initial stellar velocity distribution should be close to isotropic.   In the outskirts, outside the core, the observational situation is more complex with claims of observed tangential \citep{2019MNRAS.482.5241K,2022ApJ...939..118K}, radial \citep{2018MNRAS.480..927P,2020A&A...633A..36M}, or even mixed anisotropies depending on the axis of symmetry, the stellar component under study, or else \citep{2016MNRAS.463.1117Z,2021MNRAS.500..410L,2024ApJ...970....1V}. 

A tendency to be isotropic in the center ($\beta\sim 0$) and then radially biased ($\beta >0$) is also found in cosmological CDM zoom-in simulations of the smallest galaxies \citep{2017ApJ...835..193E,2017MNRAS.472.4786G,2023MNRAS.525.3516O}. The self-driven expansion of the \dm\ halo is not expected to occur in these simulations and so the reason has to be other than the expansion.  It is thought to be due to stellar feedback that produces episodic gas outflows, dropping the inner potential and placing stars on predominantly radial orbits \citep{2017ApJ...835..193E}. Although this sudden changes are irreversible and more effective, in a sense resemble the production of radial orbits described in Sect.~\ref{sec:example_isotropic}. Galaxy mergers may also produce radial orbits  \citep{2023MNRAS.525.3516O}.

 %
 %%%%%%%%%
 \section{Discussion}\label{sec:discussion}

The expected initial conditions described in Sect.~\ref{sec:literature} and Table~\ref{tab:literature} restrict the outcome of the mechanism in terms of the allowed stellar inner core-like structures, their extension and inner slope.  The impact of these constraints is analyzed in the present section. 

 As we argue in Sect.~\ref{sec:models},  the physical process of \dm\ expansion approximately conserves the anisotropy of the velocity distribution, at least in the innermost regions of the stellar profile. The observed dwarf galaxies seem to favor star with isotropic orbits in their center, therefore, using isotropy as initial conditions seems to be a reasonable guess. As it is shown in Appendix~\ref{sec:star_trace_dm}, even for initially isotropic orbits, the ratio between the final stellar and dark matter core sizes ($r_{c\star}/r_{c\dm}$)  is approximately the same as the original one ($r_{c0\star}/r_{0\dm}$) except for a factor of order one (see Eq.~[\ref{eq:expanssion_iso}]), 
\begin{equation}
  \frac{r_{c\star}}{r_{c\dm}} \simeq  \frac{r_{c0\star}}{r_{0\dm}}\times \left[\frac{r_{c\dm}}{r_{0\dm}}\right]^{-0.15}.
  \label{eq:expansion_iso_3}
\end{equation}
When the \dm\ expansion factor  $r_{c\dm}/r_{0\dm}$ varies from 5 to 15, the multiplying factor varies from 0.8 to 0.7. If we use the constraint that   $r_{c0\star} > 0.4\,r_{0\dm}$ (as worked out in Sect.~\ref{sec:literature} and Appendix ~\ref{app:blablabla}) then the expected initial conditions favor 
\begin{equation}
r_{c\star} > 0.30 \,r_{c\dm} ,
\label{eq:iniconst1}
\end{equation}
so that having $r_{c\star}\sim r_{c\dm}$ or even $r_{c\star} > r_{c\dm}$  cannot be discarded.

The initial \dm\ density is predicted to follow a power law with the exponent around $-c_\dm(t_0)\sim -1.5$ or a bit more negative. Since our study of the initial isotropic velocity distribution does not contemplate pure power law density profiles,  we use here the constraint derived for circular orbits (Sect.~\ref{sec:spherical}).  The stellar distribution is uncertain but is expected to be shallower than the \dm , namely,  $c_\star(t_0) < c_{\dm}(t_0)$ (Table~\ref{tab:literature}). This constraint, together with   Eq.~(\ref{eq:slopes}), implies the final stellar inner slope to be, 
\begin{equation}
  c_\star(t) < 0.6,
\end{equation}
where we have used the expected values for the initial and final \dm\ slopes; $c_\dm(t_0)=1.5$ and $c_\dm(t)=0$, respectively. Moreover, the stars will form perfect cores, $c_\star(t)= 0$, if the initial stellar slope is $c_\star(t_0)\simeq 1$ (see Eq.~[\ref{eq:constraint1}]).   

%
%%%%%
\section{Conclusions}\label{sec:conclusions}

Motivated by the observation of extended stellar cores, we explore a conceptually simple mechanism to form them in DM-dominated dwarf galaxies. A number of physical models of \dm\ beyond C\dm\ predict the thermalization of the \dm\ distribution over time, which implies the expansion of an originally small cuspy \dm\ halo to produce a core (see Sect.~\ref{sec:intro}). This expansion reduces the gravitational force that holds the stars in the halo and drives its expansion to generate diffuse stellar system.  Our analysis shows that this hypothetical mechanism works in practice. 

We have studied three types of spherically symmetric systems with different velocity anisotropies covering all extremes, from purely tangential to purely radial orbits. Action adiabatic invariants allow us to study the behavior of a stellar distribution that evolves under the effects of a slowly changing potential, generated by a distribution of dark matter with an original cusp that turns into a core. We discuss the behavior under a broad range of initial and final conditions, finding that stellar cores are produced relatively easily. However, the properties of the induced stellar cores depend on these conditions.
Thus, we reviewed the literature on the initial conditions expected from theoretical considerations, numerical simulations, and observational data, including the \dm\ and stellar density profiles, as well as the stellar velocity anisotropy.

The main results of our analysis and the subsequent numerical solution of the relevant equations are:

\noindent -
The conservation of the radial action variable implies that, for slowly changing central potentials,
 circular orbits (for which $J_r=0$) remain circular. On the other hand, the conservation of angular momentum implies that purely radial orbits (for which $L=0$) stay radial. Initially isotropic orbits remain approximately isotropic, at least in the central part of the stellar distribution (Sect.~\ref{sec:isochrone}).

\noindent - When the orbits are circular (Sect. ~\ref{sec:spherical}), initial \dm\ NFW profiles produce stellar density profiles with core-like structure  (i.e., $c_\star[t]\sim 0$) provided the initial stellar distribution has an inner slope (i.e., $c_\star[t_0]$) in the range  between 0.5 and 1.2. The stellar distribution can develop a central region lacking stars when they start off with an inner slope shallower than 0.5.  The expansion rate of the stellar distribution scales with the expansion rate of the \dm\ so that the process can also produce stellar cores smaller and larger than the \dm\ cores depending on the initial conditions.

\noindent - The mechanism acting on purely radial orbits does not produce stellar cores because radial orbits evolve into radial orbits, and the density distribution of a system of radial orbits cannot have an inner core (Sect.~\ref{sec:radial}).

\noindent - Initially isotropic orbits (Sect.~\ref{sec:isochrone}) behave very much like circular orbits. The expansion induces cores with the expansion rate of the stars being similar to that of the \dm\ driving the process.  The formed cores remains isotropic but the expansion induces the development of radially biased orbits in the outskirts of the stellar distribution. Using $\beta$ to parameterize the anisotropy (Eq.~[\ref{eq:defbeta}]), we find that it is approximately zero within the cores, with a typical value of  around 0.07 at the core radius. Outside the core, $\beta$ is positive and increases to typical values between 0.2 and 0.5 at 3 core radii (Figs.~\ref{fig:slow_expansion6} and \ref{fig:slow_expansion7}). It is qualitatively similar to the Osipkov-Merrit anisotropy. 

\noindent - Given the sensitivity of the final stellar properties on the initial conditions of the \dm\ and stars, we explore  the literature for the expected conditions suggested by theory and observations  (Sect.~\ref{sec:literature}). Table ~\ref{tab:literature} summarizes them.  The initial density of the \dm\ halos follows a power law of large negative slope ($\sim -1.5$). The initial stellar distribution is expected to be shallower than this, but with a size similar to the \dm\ distribution (see also Appendix~\ref{app:blablabla}).  Although the stellar velocities observed in the local universe are very uncertain, the centers of the dwarf galaxies tend to be isotropic.  Isotropy is likely the initial condition too provided the expansion mechanism work since it tends to conserve the type of anisotropy of the velocity distribution.

\noindent - Given the properties of the stellar distribution after expansion (Sect.~\ref{sec:models}) and the expected initial conditions (Table~\ref{tab:literature}), the stellar distribution is likely to form a core-like structure of size $r_{c\star} > 0.30 \,r_{c\dm}$, with $r_{c\dm}$ the \dm\ core of the final distribution. Thus, stellar cores with the size of the dark matter distribution or even larger can be produced by the mechanism. In addition, the final inner slope $c_\star(t)$ has to be smaller than $\sim 0.6$. Perfect stellar cores are formed if the initial stellar slope was $c_\star(t_0)\simeq 1$.

We have restricted this study to analytical models, enough to provide physical insight, simple to deal with, but difficult to compare with actual observations. Numerical modeling is needed to treat lack of spherical symmetry or to include the existence of several stellar component formed in different times and with different kinematic properties \citep[][]{2008ApJ...681L..13B,2020MNRAS.495.3022P,2023A&A...675A..49T}.
Numerical modeling is also needed to include the self-gravity of the stars, which has been neglected in our analysis. It is a good approximation for some interesting objects \citep[e.g.,][]{2024ApJ...973L..15S}, but not in general. Often baryons contribute significantly to the inner potential and then \dm\ cores can be produced by stellar feedback \citep{2010Natur.463..203G,2012MNRAS.421.3464P}. A fully self-consistent treatment of \dm\ and baryons is technically very challenging, but one can reasonably adopt the rule of thumb used in the C\dm\ framework: stellar feedback is unable to modify the \dm\ halo in galaxies with stellar mass below  $10^5$--$10^6\,M_\odot$ \citep[e.g.,][]{2012ApJ...759L..42P}. Moreover, some useful simulations exist already. We note the existence of cosmological numerical simulation of galaxy formation that considers the \dm\ to be self-interacting so that the \dm\ halos develop inner cores over time driven by   \dm\,--\,\dm\ particle collisions  \citep[SIDM; e.g.,][]{2015MNRAS.453...29E,2023MNRAS.523.4786O,2025MNRAS.536.3338C}. In these simulations, the stars form cores that  trace the evolution of the SIDM halos, in line with what we find. In particular, the size of the formed stellar cores is closely related to the size of the DM cores \citep[e.g.][]{2014MNRAS.444.3684V,2024ApJ...968L..13Z}. 

Ultimately, one would like to compare the predictions of the mechanism with observed properties of stellar cores in dwarf galaxies -- for example, the stellar and \dm\ core radii, and their dependence on stellar and \dm\ masses. This comparison requires setting up cosmological numerical simulations with various flavors of \dm\ that take self-consistently into account the baryon driven processes that reshape the distribution of \dm\  and stars. While such a direct comparison remains out of reach, the slow expansion mechanism can, in principle, generate stellar cores of any size in galaxies of arbitrary mass, given suitable initial conditions and dark matter halo expansions.
In the real galaxies, differences in initial conditions and star-formation histories are to be expected. Moreover, the \dm\ expansion factors may be very different for halos of similar mass (e.g., in SIDM,  subtle differences determine whether a particular halo expands, and so ends up with a large core, or core-collapses, and then produces a small one -- see \citeauthor{2019JCAP...12..010K}~\citeyear{2019JCAP...12..010K}; \citeauthor{2023ApJ...958L..39N}~\citeyear{2023ApJ...958L..39N}).
The mechanism likely renders a diversity of stellar cores and, in this broad sense, is consistent with observations.

\begin{acknowledgements}
Thanks are due to Claudio Dalla Vecchia for insightful discussions on various  issues addressed in the manuscript, and to Giuseppina Battaglia for comments and references on the observed $\beta$.
We also thank the anonymous referee for comments that helped clarify the arguments.
The research of JSA is partly  funded through grant PID2022-136598NB-C31 (ESTALLIDOS 8) by the Spanish Ministry of Science and Innovation (MCIN/AEI/10.13039/501100011033)  and ``ERDF A way of making Europe”.
The three authors have been supported by the European Union through the grant ``UNDARK'' of the Widening participation and spreading excellence programe (project number 101159929).
IT acknowledges support from the ACIISI, Consejer\'{i}a de Econom\'{i}a, Conocimiento y Empleo del Gobierno de Canarias and the European Regional Development Fund (ERDF) under a grant with reference PROID2021010044 and from the State Research Agency (AEI-MCINN) of the Spanish Ministry of Science and Innovation under the grant PID2022-140869NB-I00 and IAC project P/302302, financed by the Ministry of Science and Innovation, through the State Budget and by the Canary Islands Department of Economy, Knowledge, and Employment, through the Regional Budget of the Autonomous Community.
He also acknowledges support from the European Union through the grant "Excellence in Galaxies - Twinning the IAC" of the EU Horizon Europe Widening Actions  programmes (project numbers 101158446). Funding for this research was provided by the European Union (MSCA EDUCADO, GA 101119830).
Views and opinions expressed are however those of the author only and do not necessarily reflect those of the European Union or European Research Executive Agency (REA). Neither the European Union nor the granting authority can be held responsible for them.

\end{acknowledgements}

%\vspace{5mm}
%\facilities{HST (WFC\,--\,WFC3), \comment{\dots complete}}


%\software{ {\tt Scipy} \citep{2020SciPy-NMeth}}



\bibliography{biblio_paper165}{}
\bibliographystyle{aasjournal}

%
%%%%%%%%
\appendix
%%%%%%%%
%
%
%%%%
\section{General central potentials}\label{sec:general}

In the main text, we treat circular, isotropic, and radial stellar orbits as separate cases. Nevertheless, the method is versatile and can be applied to general central potentials with any anisotropy, as outlined in here.

Let as consider a general central potential $\Phi(r,p)$ that the depends on one or more
parameters, that we shall collectively denote by $p$. Because of the spherical symmetry assumption, the three action variables are $J_r$, $J_{\theta}= L -|L_z|$, and $J_{\phi} = L_z$ \citep[e.g.,][Sect.~3.5.2]{2008gady.book.....B}. There is no closed analytical expression for the action $J_r$, that  has to be evaluated numerically.  For an orbit of a particle moving in the potential
$\Phi(r,p)$ with total energy $E$, the action $J_r$ is given by the integral
\be \label{numeJr}
J_r = \frac{1}{\pi} \int_{r_{min}}^{r_{max}} \, dr \, \sqrt{2 \Big[ E-\Phi(r,p) \Bigr] - \frac{L^2}{r^2}},
\ee
where
\noindent
$r_{max}$ and $r_{min}$ are the two roots of
\be \label{rlimits}
2 \Bigl[E-\Phi(r,b) \Bigl] - \frac{L^2}{r^2} \, = \, 0.
\ee
To study the behavior of the stellar distribution when the potential changes slowly, one has first to choose a distribution in action space defined through the action variables,
\be
F(r,{\bf v},t) = \frac{1}{(2\pi)^3} {\mathcal D}(J_r,J_{\theta}+|J_{\phi}|).
\ee
\noindent
The distribution in action space stays invariant during the slow evolution of the potential from an initial potential given by parameters $p(t_0)$ to a final one given by parameters $p(t)$. To evaluate the initial phase-space density in a point $(r, {\bf v})$,  one first computes $E$ and $L$.  We note that the distribution depends only on the radial coordinate $r$, but depends on the full vector velocity ${\bf v}$ so that there is no restriction on its anisotropy. Then, though the numerical evaluation of the integral in Eq.~(\ref{numeJr}), using the initial parameters $p(t_0)$, one computes $J_r$ as a function of $r$ and ${\bf v}$. Then, keeping in mind that
\begin{equation}
 J_{\theta}+ |J_{\phi}| = L=|{\bf r}\wedge {\bf v}|,
\end{equation}
one evaluates ${\mathcal D}(J_r, J_{\theta}+|J_{\phi}|)$. This procedure gives the value of ${\mathcal D}$ at any point $(r, {\bf v})$.  
Once ${\mathcal D}(J_r, J_{\theta}+|J_{\phi}|)$ is known, one evaluates numerically the two-dimensional definite integral required to determine the initial mass density profile (Eq.~[\ref{eq:density_henon}]),
\begin{multline}
\label{inidensgen}
\rho_{\star}(r,t_0) =  4\pi \int_0^1 du \int_0^{\sqrt{2\Psi(r,t_0)}} dv \,  v^2 \,\times \\
{\mathcal D}\left(
%J_r\left[ E = \frac{v^2}{2} - \Psi(r,t_0), L = rv \sqrt{1-u^2}, p = p(t_0) \right], L = rv  \sqrt{1-u^2} \right), \\
J_r\left[ E = E(t_0), L = rv \sqrt{1-u^2}, p = p(t_0) \right], L = rv  \sqrt{1-u^2} \right), \\
%E(t_0) = \frac{v^2}{2} - \Psi(r,t_0),\\
\end{multline}
where
\begin{displaymath}
E(t) = \frac{v^2}{2} - \Psi(r,t),
\end{displaymath}
and
\begin{displaymath}
\Psi(r,t) =-\Phi[r,p(t)].
\end{displaymath}
%
%where $\Psi(r,t) =-\Phi[r,p(t)]$.
%
To clarify the meaning of the above equation, a few comments are in order. One regards $J_r$ as a function of $u$, and $v$ through $E= E(t_0)$ and $L = r v \sqrt{1-u^2}$. So does ${\mathcal D}$, which depends on $L$ in two ways, explicitly, and though $J_r$.
%
%One follows exactly
The same procedure is used to evaluate the final density profile $\rho_{\star}(r,t)$, with the same ${\mathcal D}(J_r,J_{\theta} + |J_{\phi}|)$ but using the set of parameters $p(t)$ corresponding to the final potential. That is,
\begin{multline}
\label{inidensgen}
\rho_{\star}(r,t) =  4\pi \int_0^1 du \int_0^{\sqrt{2\Psi(r,t)}} dv \,  v^2 \,\times \\
{\mathcal D}\left(
J_r\left[ E = E(t), L = rv \sqrt{1-u^2}, p = p(t) \right], L = rv  \sqrt{1-u^2} \right). 
\end{multline}

%%%%
\section{Does the final stellar core trace the underlying DM core?} \label{sec:star_trace_dm}

The short answer to the question in the title of this appendix is ``only when the original \dm\ and stellar distributions have similar sizes''. The expansion in \dm\ and stars is expected to be similar.  The argument goes as follows. Considering circular orbits, the expansion of the stars following the \dm\ core formation is controlled by Eq.~(\ref{eq:scaling}). If a perfect DM core is formed, then the final DM density in the core is equal to
\begin{displaymath}
\frac{M_\dm(<r_{c\dm},t_0)}{4\pi r_{c\dm}^3/3},
\end{displaymath}
where $r_{c\dm}$ is the DM core radius and we have assumed that the mass internal to the final \dm\ core radius has not changed much during the expansion so that
\begin{displaymath}
  M_\dm(<r_{c\dm},t)\simeq M_\dm(<r_{c\dm},t_0).
\end{displaymath}
  Moreover, because of the existence of the central core,
  \begin{displaymath}
    \left(\frac{r_{c\star}}{r_{c\dm}}\right)^3=\frac{M_\dm(<r_{c\star},t)}{M_\dm(<r_{c\dm},t)}.
  \end{displaymath}
Using these expressions, Eq.~(\ref{eq:scaling}) can be re-written as,
\begin{equation}
  \frac{r_{c\star}}{r_{c\dm}}=\left(\frac{r_{c0\star}}{r_{c\dm}}\right)^{1/4}  \left[\frac{M_\dm(<r_{c0\star},t_0)}{M_\dm(<r_{c\dm},t_0)}\right]^{1/4} ,
  \label{eq:mess1}
\end{equation}
with $r_{c0\star}$ the original radius of the stars that end at the final stellar core radius  $r_{c\star}$.  Assuming that
the original \dm\ distribution has a characteristic density with a size $r_{0\dm}$, then
\begin{equation}
\frac{M_\dm(<r_{c0\star},t_0)}{M_\dm(<r_{c\dm},t_0)}\simeq \frac{r_{c0\star}^3}{r_{0\dm}^3},
\end{equation}
which allows us to re-write Eq.~(\ref{eq:mess1}) as
\begin{equation}
  \frac{r_{c\star}}{r_{c0\star}} \simeq \left[\frac{r_{c\dm}}{r_{0\dm}}\right]^{3/4} .
  \label{eq:expanssion}
\end{equation}

Equation~(\ref{eq:expanssion}) tells that the expansion rate of the \dm\, $\sim r_{c\dm}/r_{0\dm}$, is similar to the expansion rate produced on the stars, $r_{c\star}/r_{c0\star}$. Thus, if the sizes of the original stellar and DM distribution are similar, the resulting cores are similar too. However,  the process can also produce stellar cores smaller and larger  than the \dm\ cores. It all depends on the original ratio between the stellar and \dm\ distributions. This dependence can be seen in Fig.~\ref{fig:slow_expansion2_plot}, top panel; compare the final axial ratio with original sizes parameterized by $r_{s\star}/r_{s\dm}$.

\begin{figure}[ht!] 
\centering
\includegraphics[width=0.9\linewidth]{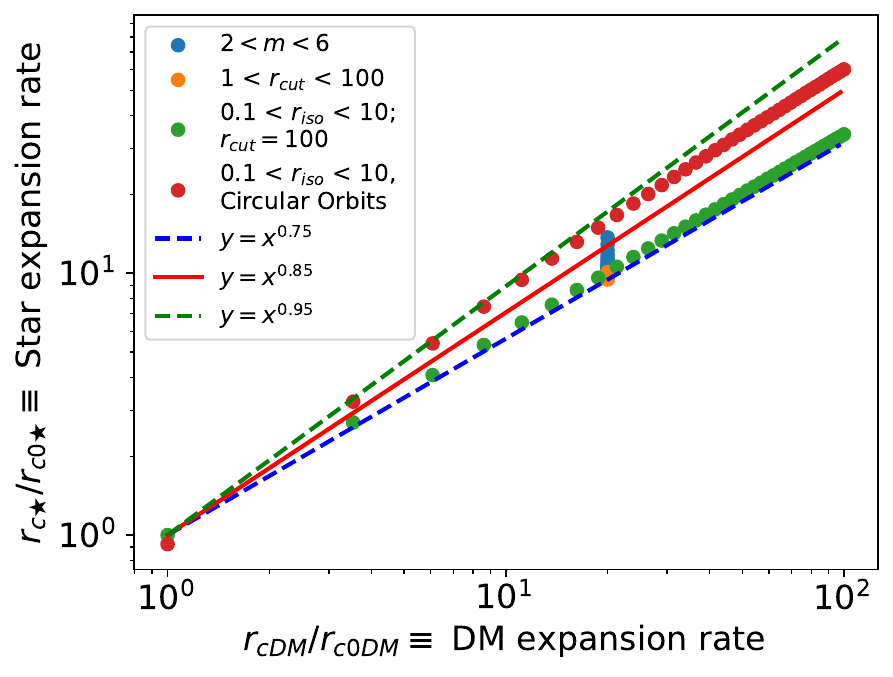}
\caption{
Scatter plot of stellar expansion rate, $r_{c\star}/r_{c0\star}$, versus the \dm\ expansion rate, $r_{c\dm}/r_{0\dm}$, for the various types of isotropic velocity distributions analyzed in Figs.~\ref{fig:slow_expansion7} and \ref{fig:slow_expansion4_plot}. The color code of the symbols and the labels are inherited from these two previous figures. The relationship is approximately a power law of exponent 0.85 (the solid red line). Two other exponents are shown for reference; in particular, 0.75 represents the scaling derived for circular orbits and given in Eq.~(\ref{eq:expanssion}).   
}
\label{fig:slow_expansion4_plot_2}
\end{figure}

The above derivation assumes circular stellar orbits, but it also holds approximately for the isotropic velocities worked out in Sect.~\ref{sec:isochrone}. The symbols in Fig.~\ref{fig:slow_expansion4_plot_2} show a scatter plot of $r_{c\star}/r_{c0\star}$ versus $r_{c\dm}/r_{0\dm}$ for the various types of potentials and expansions considered in Figs.~\ref{fig:slow_expansion7} and \ref{fig:slow_expansion4_plot}. The stellar expansion ratio $r_{c\star}/r_{c0\star}$ approximately scales with the \dm\ expansion ratio  $r_{c\dm}/r_{0\dm}$ driving the evolution. As the red solid line in Fig.~\ref{fig:slow_expansion4_plot_2} shows, the expansion is approximately given by
\begin{equation}
  \frac{r_{c\star}}{r_{c0\star}} \simeq \left[\frac{r_{c\dm}}{r_{0\dm}}\right]^{0.85} .
  \label{eq:expanssion_iso}
\end{equation}
with the exponent in between 0.75 (as in Eq.~[\ref{eq:expanssion}], represented by the dashed blue line) and 0.95 (the dashed green line).

\section{Size of the initial stellar distribution relative to the \dm }\label{app:blablabla}
\begin{figure}
\centering
\includegraphics[width=1.05\linewidth]{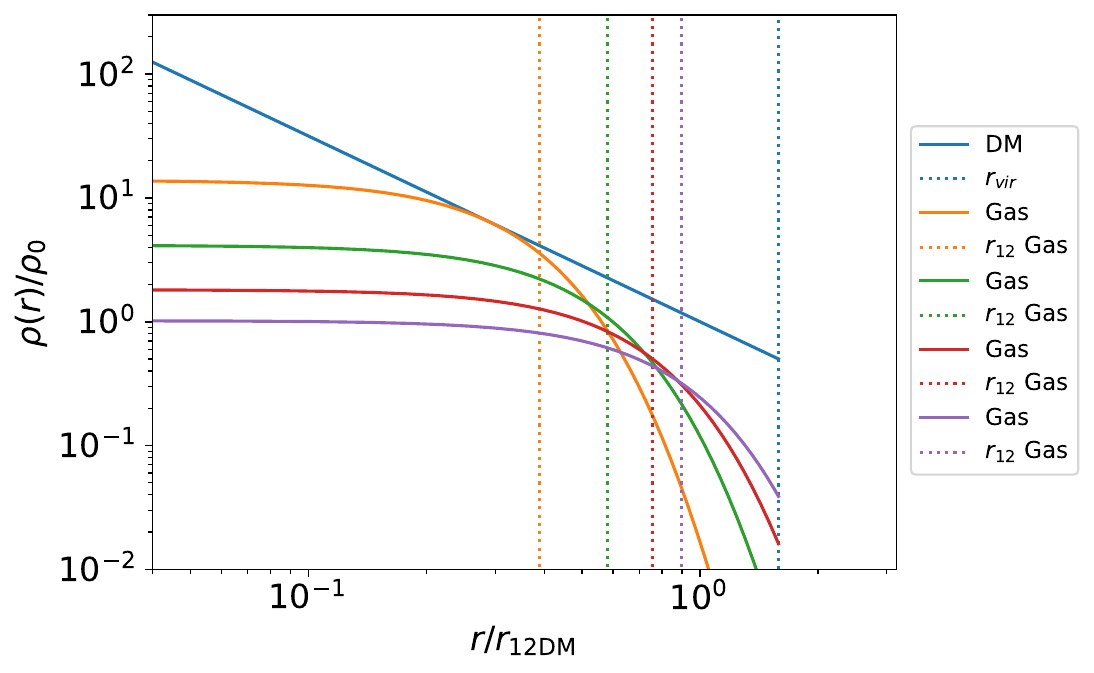}
\caption{Example of the \dm\ and gas densities used to represent the initial \dm\ halos. For the \dm\ to dominate at all radii, the sizes of the \dm\  and gas distributions have to be comparable. The power-law \dm\ density profile is shown as the blue solid line, with the virial radius marked as a blue vertical dashed line. The gas profiles are shown in different colors with the half-mass radii shown as vertical dotted lines with the color corresponding to the gas profile. Radii are normalized to the half mass radius of the \dm\ distribution $r_{12\dm}$. The half-mass radii of the gas distribution has to be larger than 0.37 times the half-mass radius for the \dm\ density to be larger than the gas density. The orange solid line shows this limiting case.}
\label{fig:blablabla}
\end{figure}
As discussed in Sect.~\ref{sec:literature}, the first \dm\ halos have a power law density distribution,
\begin{equation}
\rho_\dm(r)=
\begin{cases}
 \rho_0\, (r/r_{12\dm})^{-c_\dm}, & r \leq r_{vir},\\
  0, & \text{otherwise},
\end{cases}
\label{eq:appc1}
\end{equation}
with $r_{12\dm}$ the half-mass radius and $\rho_0=\rho_\dm(r_{12\dm})$. Independently of the halo mass (and so of the virial radius $r_{vir}$), the exponent\footnote{Following the notation in Sect.~\ref{sec:power_laws}, $c_\dm$ should formally be written as $c_\dm(t_0)$. However, for the sake of brevity, we omit the explicit $t_0$ dependence for this and other variables throughout this appendix, without any loss of clarity. }
$c_\dm$ is expected to be in the range 1.5\,--\,2.  Keeping in mind that the total mass up to a radius $r$ ($<r_{vir}$) given by Eq.~(\ref{eq:appc1}) is
\begin{equation}
M_\dm(<r)=\frac{4\pi\rho_0\,r_{12\dm}^3}{3-c_\dm}\left[\frac{r}{r_{12\dm}}\right]^{3-c_\dm},
\end{equation}
then the half-mass radius satisfies
\begin{equation}
\left(r_{12\dm}/r_{vir}\right)^{c_\dm-3}  =2.
\label{eq:appc5}
\end{equation}
The \dm\ halos forming stars for the first time collapse from material having the cosmic baryon fraction, $f_b\simeq 0.16$ \citep[e.g.,][]{2016A&A...594A..13P}. The baryons arrive as gas from which stars form and from which stars inherit proper motions and spatial distribution. As the gas radiates energy away, it cools and tends to sink toward the center of the gravitational potential, leading to a partial segregation from the \dm. The gas also experiences hydrodynamical forces, which help reaching thermodynamic equilibrium and forming a core-like feature in the central region of its distribution \citep[e.g.,][]{2008gady.book.....B}. For the sake of finding simple mathematical relations, this centrally-concentrated cored gas distribution is approximated here as a top-hat function of half-mass radius $r_{12g}$,
\begin{equation}
\rho_g(r) =
\begin{cases}
\rho_{0g}, & r \leq 2^{1/3}\,r_{12g},\\
0, & \text{otherwise},
\end{cases}
\label{eq:appc2}
\end{equation}
but more general gas distributions are used below. The total \dm\ mass for the density in Eq.~(\ref{eq:appc1}) is
\begin{equation}
  M_\dm(<r_{vir}) = \frac{8\pi\rho_0\,r_{12\dm}^3}{(3-c_\dm)},
  \label{eq:appc6}
\end{equation}
whereas the mass of the gas from Eq.~(\ref{eq:appc2}) is
\begin{equation}
M_g(<r_{vir})=\frac{8\pi}{3}\rho_{0g}r_{12g}^3,
\end{equation}
provided $2^{1/3}\,r_{12g} < r_{vir}$.
Pieced together, the two previous equations lead to
\begin{equation}
\frac{M_g(<r_{vir})}{M_\dm(<r_{vir})}=\frac{f_b}{1-f_b} = (1-c_\dm/3)\left[\frac{r_{12g}}{r_{12\dm}}\right]^3\frac{\rho_{0g}}{\rho_0}.
\label{eq:appc3}
\end{equation}
If we ask the \dm\ to be more important than the gas at all radii, as expected in \dm\ dominated  systems, this imposes
\begin{equation}
\rho_{0g} < \rho_0\,\left[\frac{2^{1/3}\,r_{12g}}{r_{12\dm}}\right]^{-c_\dm},
\end{equation}
which, together with Eqs.~Eq.~(\ref{eq:appc5}) and (\ref{eq:appc3}), lead to a lower limit for the gas half-mass radius,
\begin{equation}
  r_{12g}> r_{12\dm}\,\left[\frac{f_b}{1-f_b}\frac{2^{c_\dm/3}}{(1-c_\dm/3)}\right]^{1/(3-c_\dm)}.
\label{eq:appc7}
\end{equation}
Using typical values for the two relevant parameters ($f_b=0.16$ and $c_\dm=1.5$), one gets
\begin{equation}
r_{12g}> 0.66\, r_{12\dm},
\label{eq:appc4}
\end{equation}
a limit pretty insensitive to variation of the parameters within expected boundaries.

The limit in Eq.~(\ref{eq:appc4}) is similar to those found when, rather than a top-hat function, smooth more realistic functions are used to represent $\rho_g(r)$. For example,
for a NFW profile  (Eq.~[\ref{eq:abcprofile}], with $a,b,c=2,3,1$) one finds  $r_{12g}> 0.55\,r_{12\dm}$,
for a  Schuster-Plummer profile (Eq.~[\ref{eq:abcprofile}], with $a,b,c=2,5,0$) one finds  $r_{12g}> 0.39\, r_{12\dm}$, and for 
a profile with a inner slope different form zero (Eq.~[\ref{eq:abcprofile}], with $a,b,c=2,20,0.5$)   one finds  $r_{12g}> 0.37\, r_{12\dm}$. Figure~\ref{fig:blablabla} shows the functions used in the third example. The above values have been found numerically using Eqs.~(\ref{eq:appc5}), (\ref{eq:appc6}), and (\ref{eq:appc3}), and then determining the smallest $r_{12g}$ which stills allows $\rho_\dm(r) > \rho_g(r)$ for all $r$.
  
In short, if the initial stellar distribution inherits its extent from the gas distribution from which it was born, then the extent of the stars and the extent of \dm\ are expected to be similar (Eq.~[\ref{eq:appc4}]).


\end{document}